\documentclass[letterpaper, 10 pt, conference]{ieeeconf}  

\IEEEoverridecommandlockouts                              

\usepackage{graphicx}  
\usepackage{subcaption}
\usepackage{amsmath,bm}
\usepackage{mathtools}
\usepackage{amsfonts}
\usepackage {extarrows}
\usepackage{steinmetz}
\usepackage{siunitx}
\usepackage{amssymb}
\usepackage{mathptmx}
\usepackage{anyfontsize}
\usepackage{esvect}
\usepackage{cancel}
\usepackage{cite}
\usepackage{scalerel}

\usepackage{tikz} 
\usepackage{multirow}
\usetikzlibrary{calc,patterns,arrows,shapes.arrows,intersections}
\newtheorem{definition}{Definition}
\newtheorem{theorem}{Theorem}
\DeclarePairedDelimiter\abs{\lvert}{\rvert}%
\DeclarePairedDelimiter\norm{\lVert}{\rVert}%
\makeatletter
\let\oldabs\abs
\def\abs{\@ifstar{\oldabs}{\oldabs*}}
\let\oldnorm\norm
\def\norm{\@ifstar{\oldnorm}{\oldnorm*}}
\makeatother

\title{\LARGE \bf
Tuning of a Class of Reset Elements Using Pseudo-Sensitivities*
}
\author{Ali Ahmadi Dastjerdi$^{1}$, Niranjan Saikumar$^{1}$ and S.H. HosseinNia$^{1}$
\thanks{*This work was supported by NWO, through OTP TTW project \#16335.}
\thanks{$^{1}$A. Ahmadi Dastjerdi, N.saikumar and S.H HosseinNia are with the Faculty of Department of Precision and
    Microsystems Engineering, Delft University of Technology, Delft, The Netherlands
        {\tt\small A.AhmadiDastjerdi@tudelft.nl, N.saikumar@tudelft.nl, and\newline s.h.hosseinniakani@tudelft.nl}}%
}

\begin{document}

\maketitle
\thispagestyle{empty}
\pagestyle{empty}

\begin{abstract}
Currently, the demand for a better alternative to linear PID controllers is increasing due to the rising expectations of the high-tech industry. In literature, it has been shown that Constant in gain Lead in phase (CgLp) compensators, which are a type of reset element, have high potential to improve the performance of systems. Although there are few works which investigate tuning of these compensators, the high order harmonics and steady-state performances have not yet been considered in these methods. Recently, a frequency-domain framework has been developed to analyze closed-loop performances of reset control systems which includes high order harmonics. In this paper, this frequency-domain framework is combined with loop-shaping constraints to provide a reliable frequency-domain tuning method for CgLp compensators. Finally, different performance metrics of a CgLp compensator are compared with those of a PID controller on a precision positioning stage. The results show that the presented tuning method is effective, and the system with the CgLp compensator achieves superior dynamic performance to that of the PID controller.    
\end{abstract}
\section{INTRODUCTION}\label{sec:1}
The fast rising high-tech industry leads researchers to find a better alternative for linear controllers~\cite{dastjerdi2019linear}. One of the appropriate alternatives is reset element which has gained a lot of attention due to its simple configuration~\cite{horowitz1975non,guo2009frequency,clegg1958nonlinear,beker2004fundamental,hazeleger2016second,guo2015analysis}. In 1958, the first reset element was introduced by Clegg \cite{clegg1958nonlinear}. Clegg Integrator (CI) is an integrator which resets its state to zero when its input crosses zero. Then, First Order Reset Element (FORE)~\cite{horowitz1975non,saikumar2019constant} and Second Order Reset Element (SORE)~\cite{saikumar2019constant,hazeleger2016second} have been developed to provide more design freedom and applicability. Other reset conditions such as reset band \cite{barreiro2014reset,banos2014tuning} and fixed reset instants \cite{zheng2007improved} have also been studied. In order to soften non-linearities of reset elements, several techniques like partial reset and PI+CI approaches have been proposed \cite{banos2011reset}. 

Based on Describing Function (DF) analysis, it can be seen that reset controllers provide less lag phase in comparison with their base linear structures. This phase advantage is utilized to introduce new compensators \cite{van2018hybrid,valerio2019reset,saikumar2019constant}. One of these reset compensators is 'Constant in gain Lead in phase' (CgLp) whose gain is constant while providing a phase lead \cite{saikumar2019constant,Houu}. In these works, CgLp has been used as an alternative for the derivative to compensate part of the required phase lead. This is advantageous because the open-loop will have higher gains at low frequencies and lower gains at high frequencies which results in higher precision performances. 

There are few studies which investigate tuning of CgLp compensators~\cite{Mahmoud,saikumar2019constant,Houu}. In those works, CgLp is tuned to get a specific amount of phase lead at the cross-over frequency. However, as a result of the design flexibility of reset controllers, various combinations of tuning parameters could be used to provide the same open-loop phase compensation at the cross-over frequency based on the DF analysis. However, not all sets of tuning parameters result in performance improvement. In addition, stability has not been assessed in tuning method and has to be checked with non-linear stability methods, separately. Furthermore, the existence of the steady-state performance of the closed-loop has not been assured in those works. Thus, there is a lack of reliable tuning method for CgLp compensators. 

Recently, a new frequency-domain framework is developed which analyzes the closed-loop steady-state performances of reset control systems considering high order harmonics~\cite{ALIAUTO}. Moreover, a frequency-domain method for assessing the stability of reset elements has been proposed~\cite{AliCDC}. In this paper, we combined the frequency-domain framework~\cite{ALIAUTO}, the frequency-domain stability method~\cite{AliCDC}, the DF method, and loop-shaping constraints to provide a reliable tuning method for CgLp compensators. Finally, to show the effectiveness of the proposed tuning method, a CgLp compensator is tuned and implemented on a precision positioning stage.

In the remainder of this paper, the tuning method is elaborated in Section \ref{sec:2}. In Section \ref{sec:3}, a tuned CgLp compensator is applied to a precision positioning stage, and its performance is compared with a PID controller. Conclusions and remarks for further study are provided in Section \ref{sec:4}.
\section{Tuning Method}\label{sec:2}
In this section, first, frequency-domain descriptions for reset elements, CgLp compensators, the stability condition, and pseudo-sensitivities (sensitivity functions defined for nonlinear controllers) are briefly recalled. Then the structure of the controller is introduced, and the tuning method is proposed. 
\subsection{Frequency Analysis of Reset Elements}\label{sec:2.1}
The state-space representation of reset elements is
\begin{equation}\label{reset}
\left\{
\begin{aligned}
\dot{x}_r(t) &=A_rx_r(t)+B_re(t), & e(t)\neq0,  \\
x_r(t^+) &=A_\rho x(t), & e(t)=0, \\
u(t) &=C_rx(t)+D_re(t),
\end{aligned}
\right.
\end{equation}
in which $A_r$, $B_r$, $C_r$ and $D_r$ are the state matrices of the base linear system, $e(t)$ and $u(t)$ are the error input and control input, respectively. The resetting matrix $A_\rho$ determines states' values after reset action. Since reset elements are non-linear, the DF analysis is popularly used in literature to study their frequency behaviour. To have a well-defined steady-state response, it is required that $A_r$ has all eigenvalues with negative real part and $A_\rho e^{\frac{A_r\pi}{\omega}}$ has all eigenvalues with magnitude smaller than one \cite{guo2009frequency}. The sinusoidal input DF of reset elements (\ref{reset}) is given in \cite{guo2009frequency} as
 \begin{equation}
 {\mathcal{N}( j \omega ) = C_r  \left( j \omega I - A_r \right) ^ { - 1 } B_r  \left( I + j \Theta ( \omega ) \right) + D_r },
\end{equation}
 where $\Theta $ is
\setlength{\arraycolsep}{0.0em}
\begin{eqnarray}\label{E-02-03-01}
\Theta(\omega)&{=}&\frac{-2\omega^2}{\pi}(I+e^{\frac{\pi A_r}{\omega}})\Bigg((I+A_\rho e^{\frac{\pi A_r}{\omega}})^{-1}A_\rho(I+e^{\frac{\pi A_r}{\omega}})\nonumber\\
&&{-}\:I\Bigg)(\omega^2I+A_r^2)^{-1}.
\end{eqnarray}
\setlength{\arraycolsep}{5pt}\subsection{CgLp Compensator}\label{sec:2.2}
\begin{figure}[t!]
    \centering
    \begin{tikzpicture}
    \node[anchor=south west,inner sep=0] at (0,0) {\includegraphics[width=\linewidth]{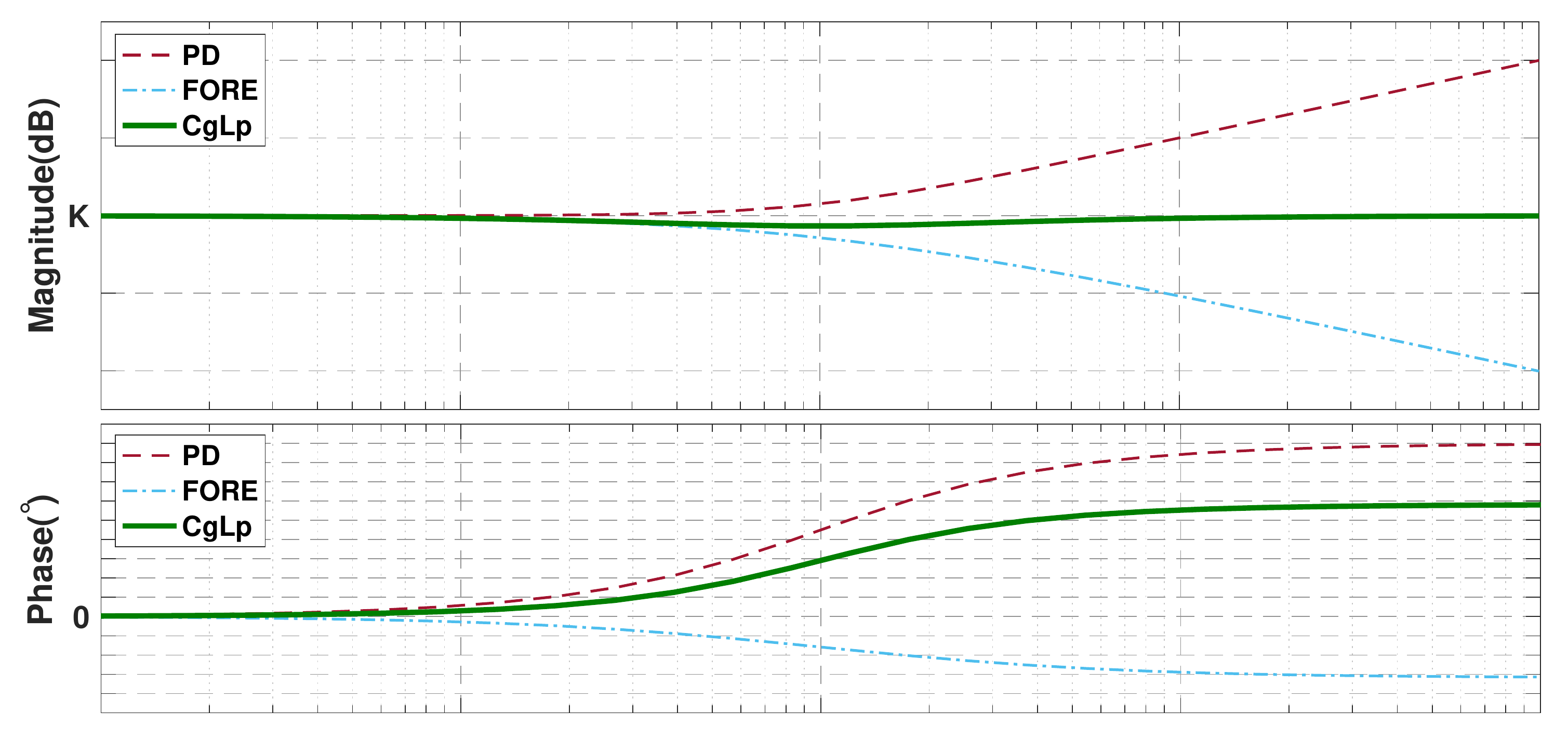}};
\draw (4.6,0) node [scale=0.6]  {$\bm{\omega_r=\omega_d}$};
\draw (0.48,1.3) node [scale=0.6]  {$\bm{\theta}$};
\end{tikzpicture}
   \caption{The DF of a CgLp compensator}  
  \label{F-01}
\end{figure}

A CgLp compensator (\ref{E-01}) is constructed using a FORE or a SORE with the series combination of a corresponding order of a lead filter. Considering the DF analysis, this compensator has a constant gain with a lead phase (Fig.~\ref{F-01}) (\cite{saikumar2019constant,saikumar2019complex}). In this paper, we only consider the first order CgLp which is 
\begin{equation}\label{E-01}
C_{CgLp}(s)=\left(\cancelto{\gamma}{\frac{1}{\frac{s}{\omega_r}+1}}\right)\left(\frac{\frac{s}{\omega_d}+1}{\frac{s}{\omega_t}+1}\right),
\end{equation}
where $\omega_r$ is the corner frequency of the reset element, $A_\rho=\gamma$ is the reset matrix, and  $\omega_d$ and $\omega_f$ are the corner frequencies of the lead filter. To have a constant gain, corner frequencies $\omega_d$ and $\omega_r$ are almost equal (there is a small correction factor which is provided in~\cite{saikumar2019constant}) and $\omega_f\gg\omega_r$.  
\subsection{$H_\beta$ Condition}\label{sec:2.3}
There are several theories to determine the stability of reset control systems~\cite{banos2011reset,guo2015analysis,beker2004fundamental,polenkova2012stability}. Among those, the $H_\beta$ condition presented in~\cite{beker2004fundamental,guo2015analysis,AliCDC} gets a lot of attention. In~\cite{AliCDC}, a method is developed to examine the $H_\beta$ conditions utilizing the frequency response of the plant. To this end, consider $L(j\omega)$ and $C_R(j\omega)$ as the base linear frequency responses of the open-loop and of the reset element, respectively. Then, the Nyquist Stability Vector (NSV=$\vv{\mathcal{N}}(\omega)\in\mathbb{R}^2$), for all $\omega\in\mathbb{R}^+$, is $\vv{\mathcal{N}}(\omega)=[\mathcal{N}_\chi \quad \mathcal{N}_\Upsilon]^T$ in which
\setlength{\arraycolsep}{0.0em}
\begin{eqnarray}
\mathcal{N}_\chi&{=}&\abs{L(j\omega)+\frac{1}{2}}^2-\frac{1}{4},\nonumber\\
\mathcal{N}_\Upsilon&{=}&\Re(L(j\omega)\cdot C_R(j\omega))+\Re(C_R(j\omega)).\nonumber
\end{eqnarray}
\setlength{\arraycolsep}{5pt}\begin{theorem}\label{T1}
Considering $\theta_{1}=\underset{\omega\in\mathbb{R}^+}{\min}\phase{\vv{\mathcal{N}}(\omega)}$ and $\theta_{2}=\underset{\omega\in\mathbb{R}^+}{\max}\phase{\vv{\mathcal{N}}(\omega)}$. Suppose $-1<\gamma\leq1$, then, the $H_\beta$ condition for a reset control system is satisfied and its response is uniformly bounded-input bounded-state (UBIBS) for any bounded input if~\cite{AliCDC} 
\begin{equation}\label{E-3333}
\left(-\dfrac{\pi}{2}<\theta_{1}<\pi\right)\ \land\ \left(-\dfrac{\pi}{2}<\theta_{2}<\pi\right)\ \land\ (\theta_{2}-\theta_{1}<\pi).
\end{equation}	
\end{theorem}
\subsection{Pseudo-Sensitivities for Reset Control Systems}\label{sec:2.4}
In linear systems, the relation between reference signal $r(t)$ to error $e(t)$ can be calculated by sensitivity transfer functions~\cite{schmidt2014design}. Although it is possible to use the DF of the reset elements in those sensitivity transfer functions to analyze the tracking performance of CgLp compensators, it is not a reliable approach because high order harmonics are neglected. In order to analyze reset control systems more accurately, a pseudo-sensitivity function $S_{\infty}(j\omega)$ for a sinusoidal reference $r=r_0\sin(\omega t)$ is defined in \cite{ALIAUTO}. 
\begin{theorem}\label{T2}
A closed-loop reset control system has a well-defined steady-state solution for any Bohl function input if the $H_\beta$ condition is satisfied and reset instants have the well-posedness property~\cite{ALIAUTO}.
\end{theorem}
In addition, if Theorem~\ref{T2} holds, the tracking error of the reset control system is a periodic function with the period $\dfrac{2\pi}{\omega}$. Thus, the pseudo-sensitivity for a reset control system is defined as the ratio of the maximum tracking error of the system to the magnitude of the reference at each frequency. 
\begin{definition}{Pseudo-sensitivity $S_\infty$}\label{d1}
$$
\begin{array}{*{35}{c}}
\forall \omega\in\mathbb{R}^{+}:\ S_\infty(j\omega)=e_{\max}(\omega)e^{j\varphi_{max}},
\end{array}
$$ 
\end{definition} 
where 
$$e_{\max}(\omega)=\left(\dfrac{\underset{t_{ss_0}\leq t\leq t_{ss_m}}{\max}(r(t)-y(t))}{|r|}\right)=\sin(\omega t_{\max})-\dfrac{y(t_{\max})}{r_0},$$
$\varphi_{max}=\frac{\pi}{2}-\omega t_{max}$, $y(t)$ is the response of the closed-loop reset control system, and $t_{ss_0}$ and $t_{ss_m}=t_{ss_0}+\frac{2\pi}{\omega}$ are the steady-state reset instants of the closed-loop reset control system $(e(t_{ss_0})=e(t_{ss_m})=0)$. In a similar way, the pseudo-control sensitivity $CS_\infty(\omega)$, the pseudo-complementary sensitivity $T_\infty(\omega)$, and the pseudo-process sensitivity $PS_\infty(\omega)$ are defined in \cite{ALIAUTO}. These calculations are embedded in a user-friendly toolbox~\cite{Toolbox}.  
\subsection{Problem Formulation}\label{sec:2.5}
In this section, the tuning procedure is explained. For this purpose, a CgLp compensator along with a PID controller is considered for tuning as
\begin{equation}\label{E-5}
C _ {\text{CgLp}}(s) = K_p\underbrace{\left(\cancelto{\gamma}{\dfrac{1}{\frac{\alpha s}{\omega_r}+1}}\right)\left(\dfrac{\frac{s}{\omega_r}+1}{\frac{s}{\omega_f}+1}\right)}_{\mathrm{CgLp}}\underbrace{\overbrace{\left(1+\frac{\omega_i}{s}\right)}^{\mathrm{PI}}\overbrace{\left(\dfrac{\frac{s}{\omega_d}+1}{\frac{s}{\omega_t}+1}\right)}^{\mathrm{Lead}}}_{\mathrm{PID}},
\end{equation}
in which $\gamma=A_\rho$ determines the value of the reset state after the reset action and $(K_p,\omega_i,\omega_r,\omega_t,\omega_d,\omega_f,\gamma)$ is the tuning parameter set. It has been shown that the sequence of controller filters has effects on the performance of reset control systems~\cite{Caipaper}. In this research, the sequence of control filters is the traditional approach in which the tracking error is the input of the reset element and other linear parts following in series. In this tuning method, the controller is tuned given the following specifications: cross-over frequency $\omega_c$, phase margin $\varphi_m$, and modulus margin $M_m$. Note that these specifications are based on the DF analysis or defined pseudo-sensitivities. Furthermore, since the scope of this paper is tuning of the CgLp part, $\omega_i$ and $\omega_f$ are tuned as $\dfrac{\omega_c}{10}$ and $8\omega_c$, respectively, to have acceptable tracking and noise rejection performances~\cite{schmidt2014design,dastjerdi2018tuning,saikumar2019constant}. To assure stability and use of pseudo-sensitivities, the $H_\beta$ condition (Theorem~\ref{T1}) has to be satisfied. In addition, a robustness requirement in the form of iso-damping behaviour~\cite{dastjerdi2019linear,de2016novel} requires that the phase behaviour of the system must follow
\begin{equation}\label{ER1}
\dfrac{d(\phase{\mathcal{N}_{\text{CgLp}}(j\omega)\text{PID}(j\omega)G(j\omega)})}{d\omega}\Big|_{\omega=\omega_c}=0.
\end{equation}
All constraints are summarized as
\begin{itemize}
\item Cross-over frequency constraint: $$|\mathcal{N}_{\text{CgLp}}(j\omega_c)\text{PID}(j\omega_c)G(j\omega_c)|=1$$
\item Phase margin constraint: $$\phase{\mathrm{\mathcal{N}_{CgLp}}(\omega_c)}+\phase{\mathrm{PID}(\omega_c)}+\phase{G(\omega_c)}=\varphi_m$$
\item Modulus margin constraint: $\text{max }|S_\infty(j\omega)|<M_m$
\item Iso-damping Behaviour: Equation (\ref{ER1})
\item The $H_\beta$ condition: Equation (\ref{E-3333}) and $-1<\gamma\leq1$ 
\end{itemize}
Eventually, we define a suitable cost function for the tuning of the control structure~(\ref{E-5}). According to~\cite{sabatier2015fractional}, to have an appropriate tracking performance in the interested region of frequencies, the following cost function is obtained as
\begin{equation}\label{E-306}
J=\underset{\omega\leq\omega_l}{\max}\Big|\dfrac{S_\infty(j\omega)}{\omega}\Big|_{\text{dB}},
\end{equation}   
in which $\omega_l$ determines the interested region of frequencies over which the reset control system is expected to track references and reject disturbances. There are several methods such as grid search, gradient methods, Genetic Algorithm, etc., for solving this problem. Here, since the performance of the controller is not so sensitive to a small change of the tuning set parameter, we suggest to use a grid search method for completing the tuning procedure. The parameter $K_p$ is determined by the cross-over frequency definition. In addition, it is possible to find vectors $l_B$ and $u_B$ to set lower and higher limits for $(\omega_r\, \omega_d\, \omega_t)$ by the phase margin definition and considering stability of the base linear of the system (i.e. $\omega_c l_B<[\omega_r\ \omega_d\ \omega_t]^T<\omega_c u_B$). Then, with a small resolution, we grid the parameters and provide a parameter space. Now, we calculate constraints (3)-(6) for every point in this space, and eliminate the points which do not satisfy the constraints. Finally, suppose there are N tuning parameter sets $(K_p,\omega_r,\omega_t,\omega_d,\gamma)$ which satisfy the aforementioned constraints, then the parameter set which has the minimum $J$ value is selected for designing the control structure~(\ref{E-5}).  
\section{Practical Example}\label{sec:3}
\begin{figure}[!t]
\centering
\includegraphics[scale=0.45]{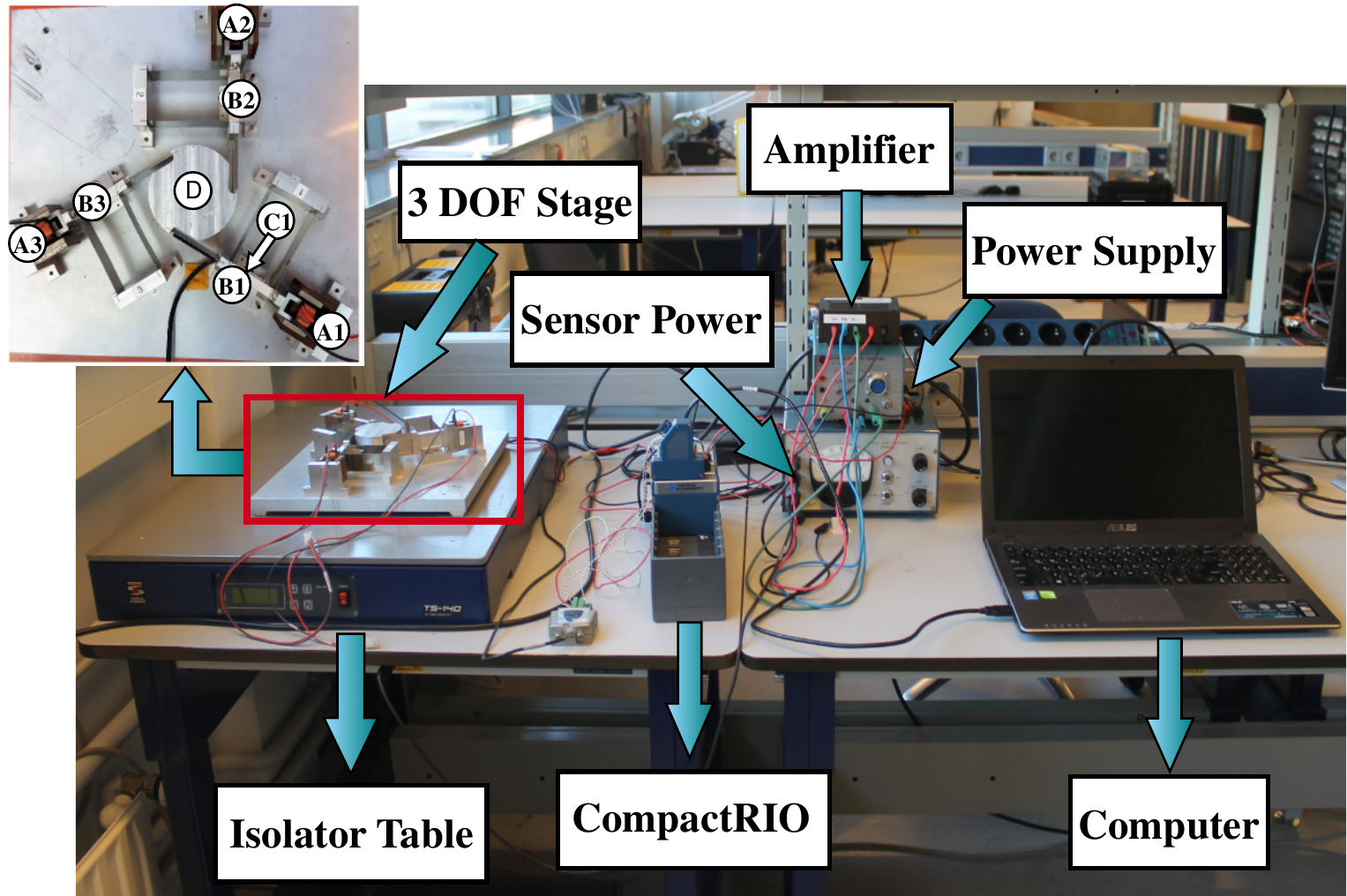}
\caption{The whole setup including computer, CompactRio, power supply, sensor power, amplifier, isolator, sensor and, stage}
\label{F-3}
\end{figure}

\begin{figure}[!t]
\centering
\includegraphics[scale=0.28]{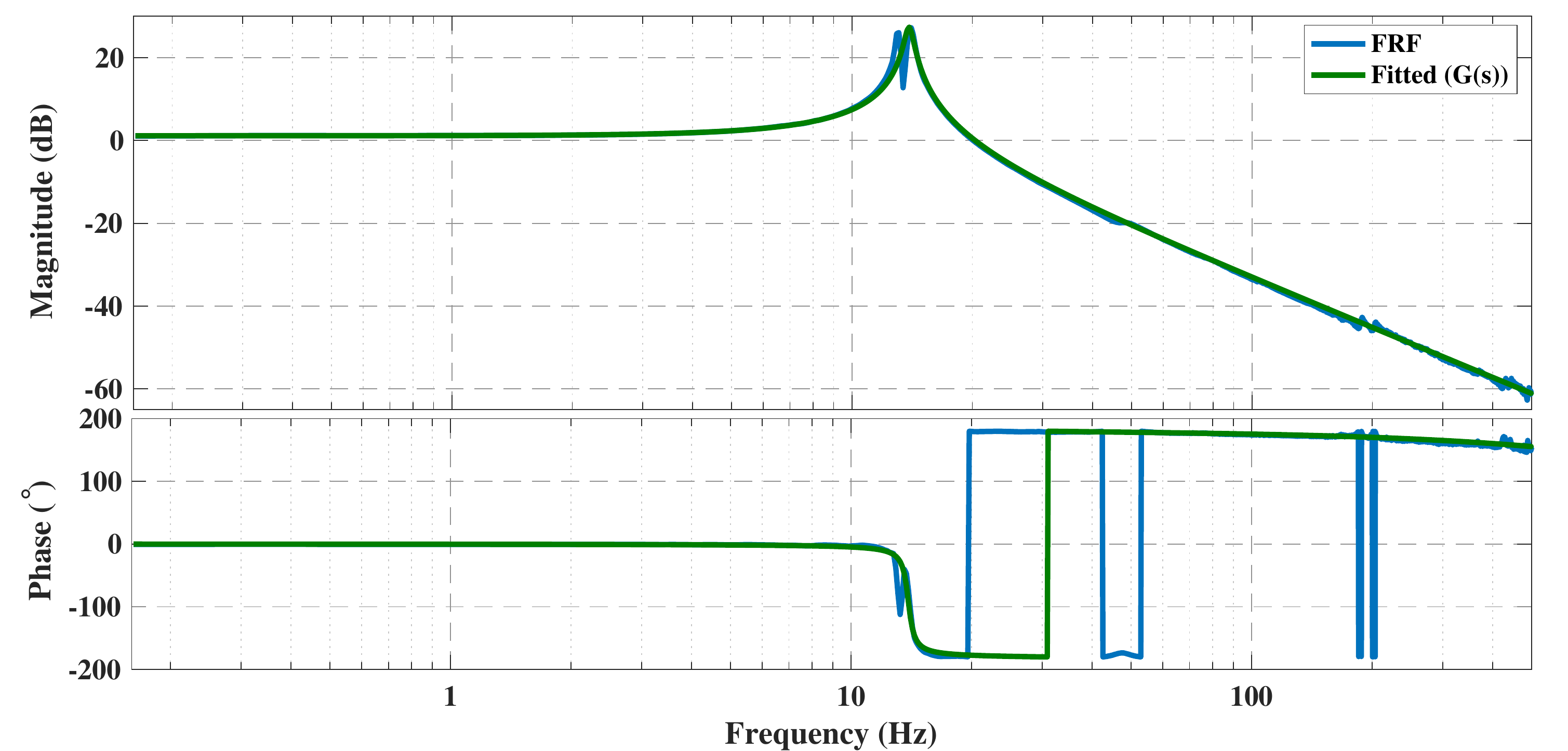}
\caption{Identification of the stage}
\label{F-4}
\end{figure}

\begin{figure}[thpb]
\centering
\includegraphics[width=\linewidth]{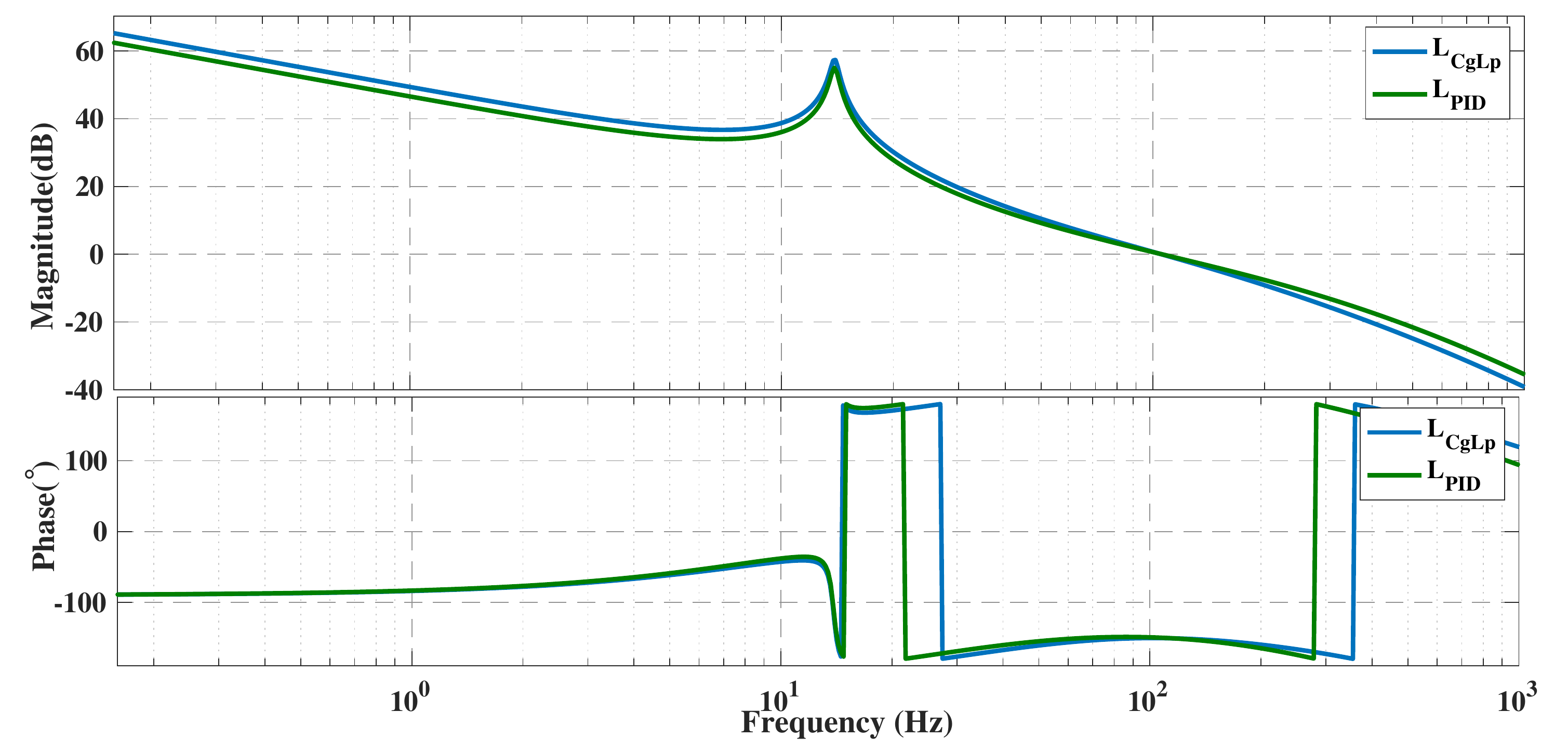}
\caption{Open-loop frequency responses of the system with controllers $C_{{\text{PID}}}$ and $C_{\text{CgLp}}$}  
  \label{F-05}
\end{figure}

To show the effectiveness of the proposed tuning method, a precision positioning stage (Fig~\ref{F-3}) is used as a benchmark in this paper. In this stage, which is termed ``Spider", three actuators are angularly spaced to actuate 3 masses (indicated by B1, B2, and B3) which are constrained by parallel flexures and connected to the central mass D through leaf flexures. Only one of the actuators (A1) is considered and used for controlling the position of mass B1 attached to the same actuator which results in a SISO system. A linear power amplifier is utilized to drive the Lorentz actuator, and Mercury M2000 linear encoder is used to obtain position feedback with the resolution of \SI{0.1}{\micro\meter}. The identified frequency response data of the system is shown in Fig~\ref{F-4}. As shown in Fig.~\ref{F-4}, although the plant is a collocated double mass-spring system, the identified frequency response data is well approximated by a mass-spring-damper system with the transfer function
\begin{equation}\label{E-47}
 G(s)\approx\frac{Ke^{-\tau s}}{\frac{s^2}{\omega_r^2}+\frac{2\zeta s}{\omega_r}+1}=\frac{1.14e^{-0.00014s}}{\frac{s^2}{7627}+\frac{0.05s}{87.3}+1}.
\end{equation} 
 Note that to use relations provided in \cite{ALIAUTO}, the time delay $(e^{-0.00014s})$ is approximated by the first order Pade method \cite{al2000approximation} as $\dfrac{-s+14400}{s+14400}$. The design requirements for this system are: 
\begin{itemize}
\item the cross-over frequency: $\omega_c=100 \text{ Hz}$ 
\item the phase margin: $\varphi_m=30^{\circ}$
\item the modulus margin: $M_m \leq 6.5\text{ dB}$
\end{itemize} 
Now, the control structure~(\ref{E-5}) is tuned based on the described method in Section~\ref{sec:2.5}. To satisfy constraint~(\ref{ER1}), $\omega_c\leq\omega_t$. Also, in order to not eliminate the low-pass filter effects, $\omega_t\leq\omega_f$. Considering $30^{\circ}$ phase margin and stability of the base linear system, it is obtained that $\dfrac{\omega_c}{5}<\omega_d<\omega_c$. Furthermore, selecting a very small value for $\omega_r$ leads to increase the amplitude of high order harmonics at low frequencies which are not desired~\cite{Houu}. Hence, we consider this parameter range $\omega_c[0.05\ 0.2\ 1]^T<[\omega_r\ \omega_d\ \omega_t]^T<[1\ 1\ 8]^T\omega_c$ in the tuning procedure. In addition, we take $\omega_l=\dfrac{\omega_c}{10}$ as the maximum limit of the interest region for tracking. The controller is obtained through the proposed tuning method as	
\begin{equation}\label{E-41}
\resizebox{\hsize}{!}{$
C_{\text{CgLp}}{=}25.5\left(\cancelto{0.3}{\dfrac{1}{\dfrac{s}{111\pi}+1}}\right)\left(\dfrac{\dfrac{s}{105.2\pi}+1}{\dfrac{s}{1600\pi}+1}\right)\left(1+\dfrac{20\pi}{s}\right)\left(\dfrac{\dfrac{s}{105.2\pi}+1}{\dfrac{s}{260\pi}+1}\right).$}
\end{equation} 
To compare the performance of the tuned controller with a linear controller, a PID structure is also tuned with the same method proposed in Section~\ref{sec:2.5}. To have a fair comparison, the structure of the PID controller is similar to the control structure~(\ref{E-5}). Finally, the $C_{\text{PID}}$ is  
\begin{equation}\label{E-42}
C_{\text{PID}}=18.46\left(\dfrac{1}{\dfrac{s}{1600\pi}+1}\right)\left(\dfrac{\dfrac{s}{77\pi}+1}{\dfrac{s}{520\pi}+1}\right)\left(1+\frac{20\pi}{s}\right).
\end{equation}  
Figure \ref{F-05} shows the open-loop frequency response of the system with controllers $C_{\text{PID}}$ and the DF of the open-loop of the system with the controller $C_{\text{CgLp}}$.  Two systems have the same phase margin and are robust against the gain variation (iso-damping behaviour) as shown in Fig.~\ref{F-05}.   
\begin{figure*}
	\centering
	\begin{subfigure}[b]{0.49\columnwidth}
		\centering
		\includegraphics[width=\hsize]{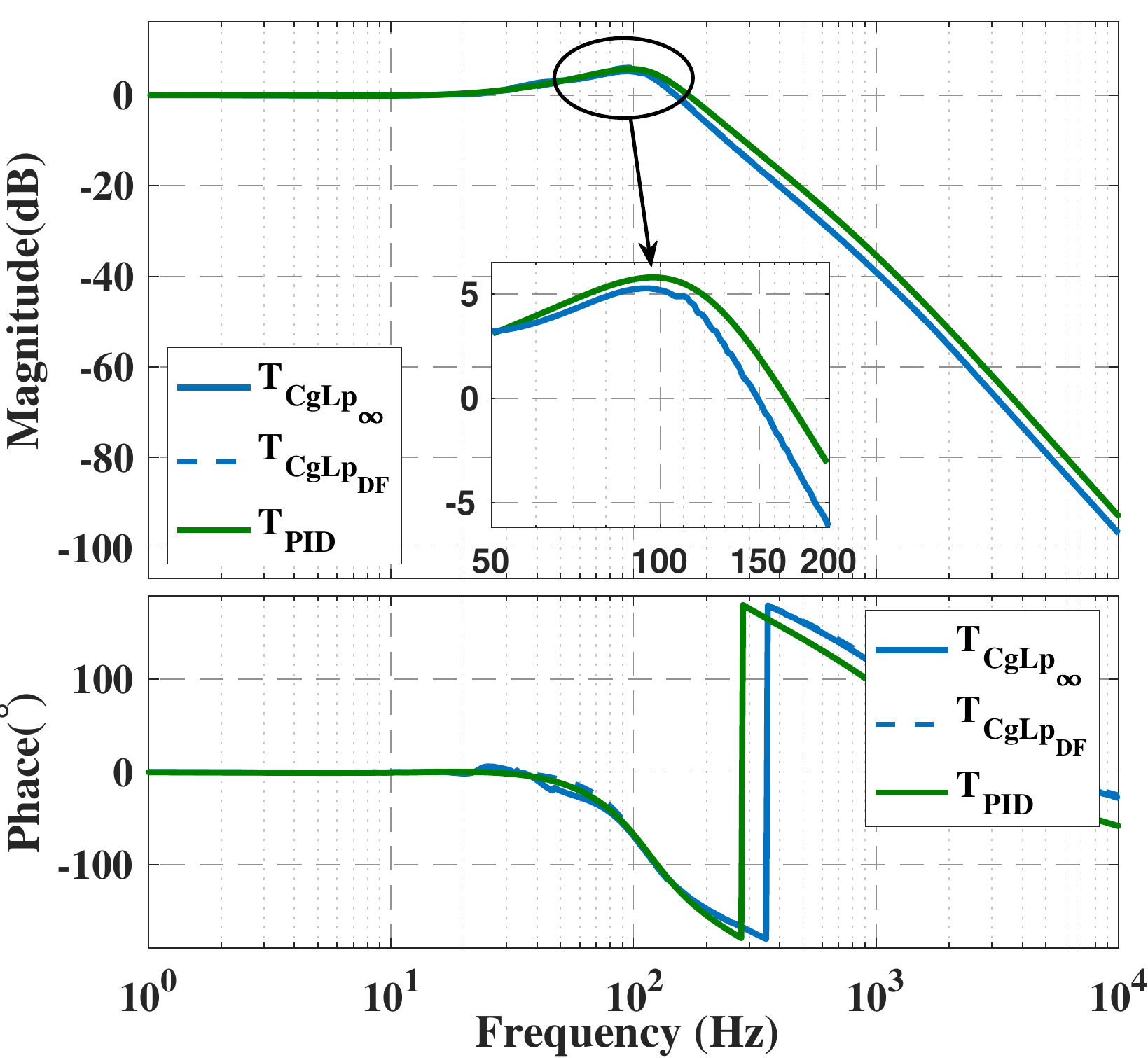}
		\caption{Complementary sensitivity}
		\label{F-06a}
	\end{subfigure}
	\hfil
	\begin{subfigure}[b]{0.49\columnwidth}
		\centering    
		\includegraphics[width=\hsize]{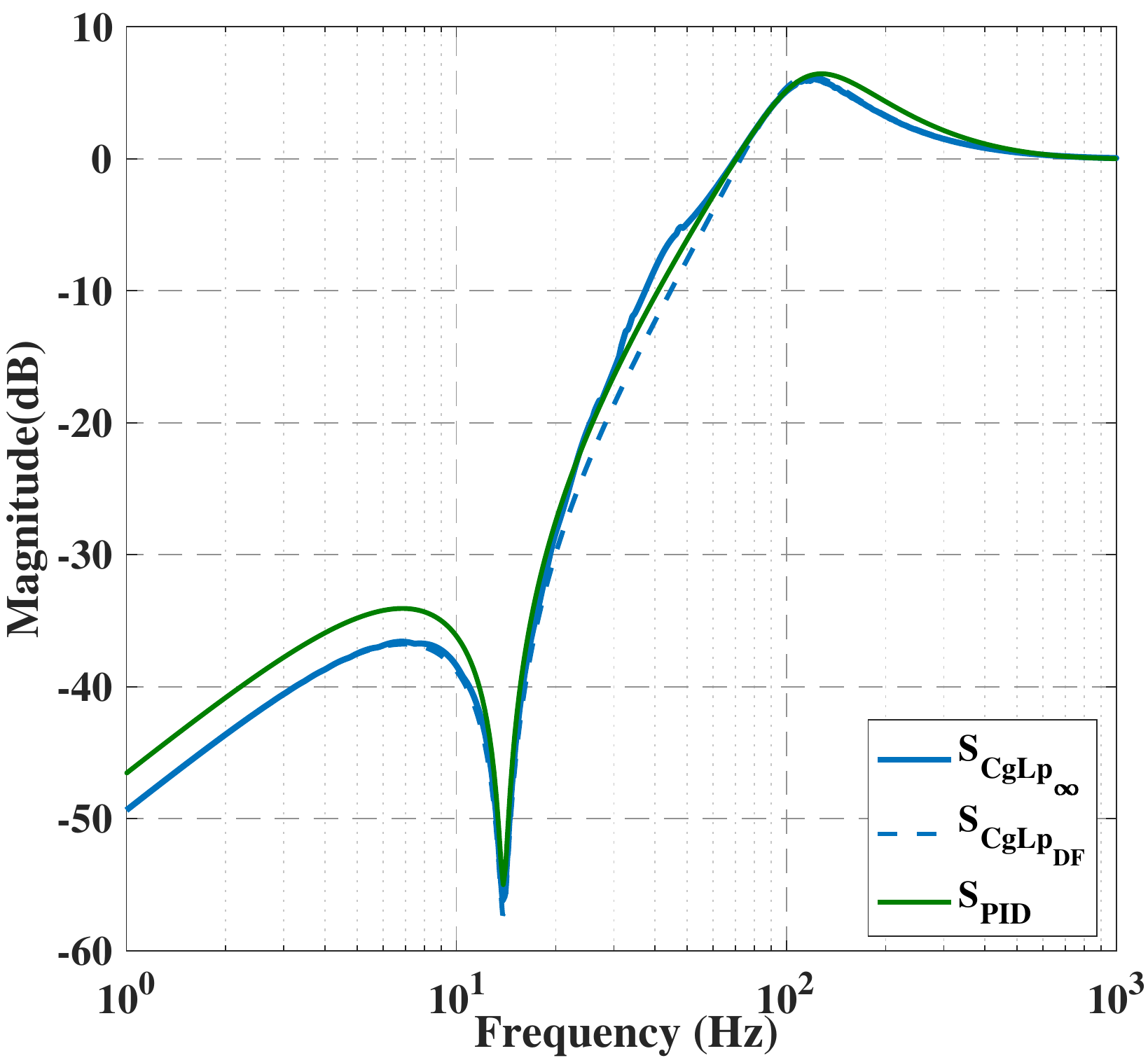}
		\caption{Sensitivity}        
		\label{F-06b}
	\end{subfigure}
	\hfil
	\begin{subfigure}[b]{0.49\columnwidth}
		\centering
		\includegraphics[width=\hsize]{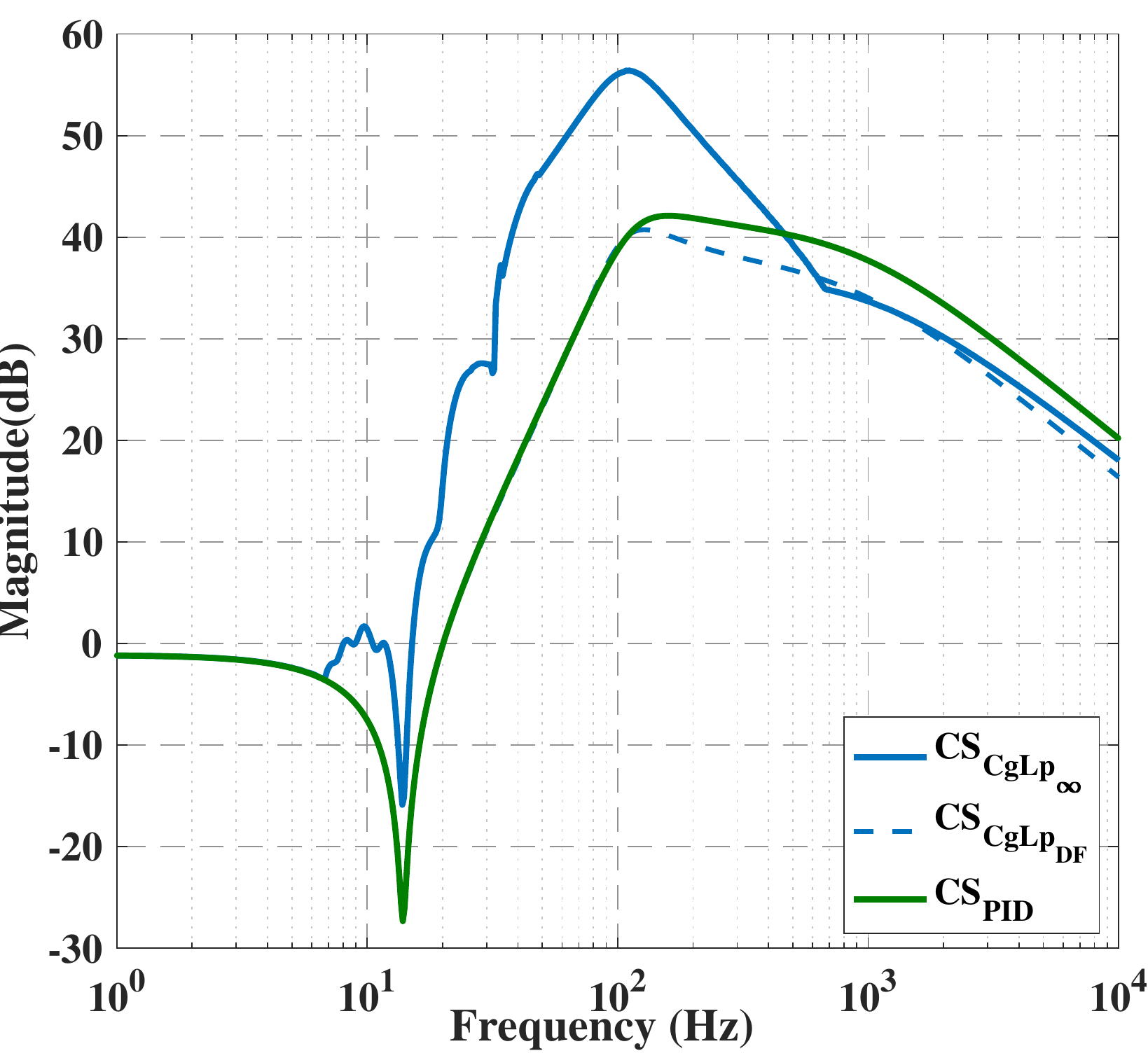}
		\caption{Control sensitivity}
		\label{F-07b}
	\end{subfigure}
	\hfil
	\begin{subfigure}[b]{0.49\columnwidth}
		\centering 
		\includegraphics[width=\hsize]{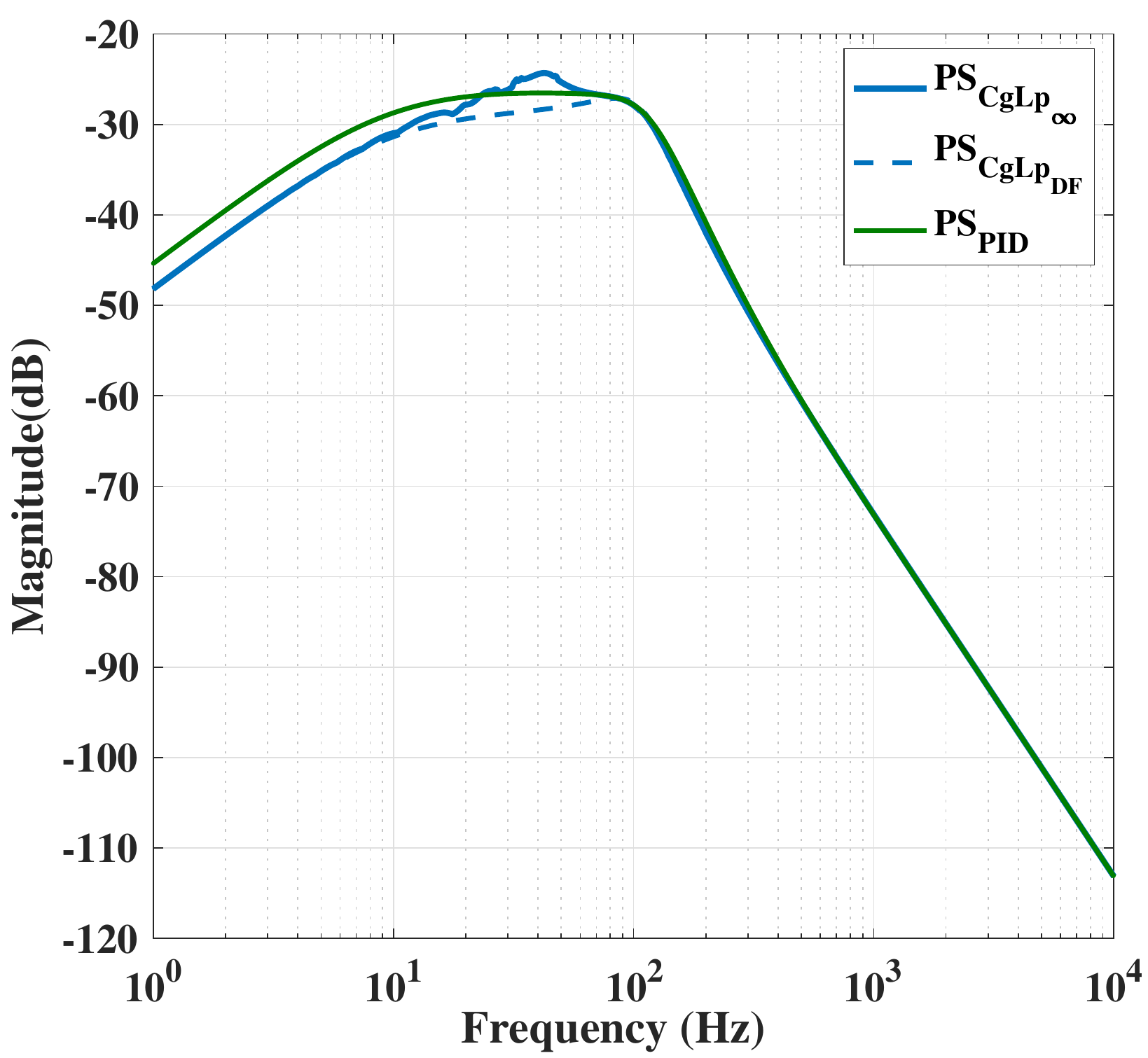}
		\caption{Process sensitivity}
		\label{F-07a}
	\end{subfigure}
	\caption{The DFs $(.\_$ DF) and pseudo-sensitivities (.$\_\ \infty$) of the closed-loop of the system with the controllers $C_{CgLp}$, and the closed-loop sensitivities of the system with the controller $C_{\text{PID}}$}
	\label{F-07}
\end{figure*}

The closed-loop frequency responses of the systems with the controller $C_{\text{CgLp}}$ including the pseudo-sensitivities and the DF methods, and the closed-loop sensitivities of the system with the controller $C_{{\text{PID}}}$ are shown in Fig.~\ref{F-07}. These frequency responses are obtained utilizing the toolbox in~\cite{Toolbox}. 
By $T_\infty$ (Fig.~\ref{F-06a}), the noise rejection capability of the system with the controller $C_{\text{CgLp}}$ must be better than that of the controller $C_{\text{PID}}$. Furthermore, as shown in Fig.~\ref{F-06b}, the system with the controller $C_{\text{CgLp}}$ has better tracking performance than that one with the controller $C_{{\text{PID}}}$ at frequencies less than $10$Hz while the modulus margin of the system with the controller $C_{\text{CgLp}}$ is less than that of with the controller $C_{\text{PID}}$. Also, there are discrepancies between the sensitivity DF and pseudo-sensitivity in the frequency range (30 - 70 Hz) which are due to the existence of high order harmonics. 
\newline Based on $PS_\infty$ (Fig~\ref{F-07a}), the disturbance rejection capability of the system with the controller $C_{\text{CgLp}}$ is better than that of the controller $C_{{\text{PID}}}$. As shown in~Fig.\ref{F-07b}, there is a significant difference between the control input of the system with the controller $C_{\text{CgLp}}$ and what is predicted by the DF method. In addition, the control input of the system with the controller $C_{\text{CgLp}}$ is more than one with the controller $C_{{\text{PID}}}$. This is explained by the fact that reset elements produce jumps in their output and differentiation of jumps produces a large control input.    
\subsection{Time Domain Results}\label{sec:3.2}
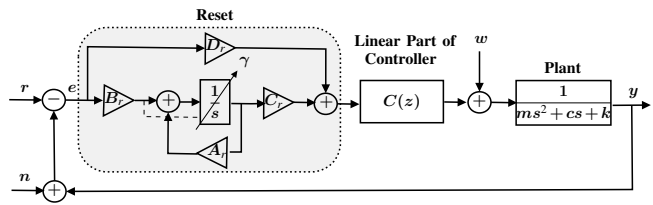
\begin{figure}[thbp]
  \centering
\resizebox{\hsize}{!}{
\tikzset{every picture/.style={line width=0.75pt}} 
\begin{tikzpicture}[x=0.75pt,y=0.75pt,yscale=-1,xscale=1]
\draw  [line width=1.5]  (462.5,100) -- (570.5,100) -- (570.5,155) -- (462.5,155) -- cycle ;
\draw  [line width=1.5]  (48.63,125.45) .. controls (48.63,117.84) and (55.3,111.68) .. (63.53,111.68) .. controls (71.76,111.68) and (78.43,117.84) .. (78.43,125.45) .. controls (78.43,133.06) and (71.76,139.22) .. (63.53,139.22) .. controls (55.3,139.22) and (48.63,133.06) .. (48.63,125.45) -- cycle ;
\draw [line width=1.5]    (815.63,132.18) -- (816.5,242) -- (83.43,244.44) ;
\draw [shift={(79.43,244.45)}, rotate = 359.81] [fill={rgb, 255:red, 0; green, 0; blue, 0 }  ][line width=0.08]  [draw opacity=0] (11.61,-5.58) -- (0,0) -- (11.61,5.58) -- cycle    ;
\draw [line width=1.5]    (5.5,125) -- (44.63,125.41) ;
\draw [shift={(48.63,125.45)}, rotate = 180.6] [fill={rgb, 255:red, 0; green, 0; blue, 0 }  ][line width=0.08]  [draw opacity=0] (11.61,-5.58) -- (0,0) -- (11.61,5.58) -- cycle    ;
\draw [line width=1.5]    (790.77,132.37) -- (836.5,132.03) ;
\draw [shift={(840.5,132)}, rotate = 539.5799999999999] [fill={rgb, 255:red, 0; green, 0; blue, 0 }  ][line width=0.08]  [draw opacity=0] (11.61,-5.58) -- (0,0) -- (11.61,5.58) -- cycle    ;
\draw [line width=1.5]    (375.65,128.35) -- (400.66,128.78) ;
\draw [shift={(404.66,128.85)}, rotate = 180.99] [fill={rgb, 255:red, 0; green, 0; blue, 0 }  ][line width=0.08]  [draw opacity=0] (11.61,-5.58) -- (0,0) -- (11.61,5.58) -- cycle    ;
\draw [line width=1.5]    (571.62,127.95) -- (596.63,128.38) ;
\draw [shift={(600.63,128.45)}, rotate = 180.99] [fill={rgb, 255:red, 0; green, 0; blue, 0 }  ][line width=0.08]  [draw opacity=0] (11.61,-5.58) -- (0,0) -- (11.61,5.58) -- cycle    ;
\draw  [line width=1.5]  (665.17,100.09) -- (788.5,100.09) -- (788.5,158) -- (665.17,158) -- cycle ;
\draw  [fill={rgb, 255:red, 241; green, 241; blue, 241 }  ,fill opacity=1 ][dash pattern={on 1.69pt off 2.76pt}][line width=1.5]  (95.5,68.6) .. controls (95.5,47.83) and (112.33,31) .. (133.1,31) -- (398.9,31) .. controls (419.67,31) and (436.5,47.83) .. (436.5,68.6) -- (436.5,181.4) .. controls (436.5,202.17) and (419.67,219) .. (398.9,219) -- (133.1,219) .. controls (112.33,219) and (95.5,202.17) .. (95.5,181.4) -- cycle ;
\draw [line width=1.5]    (307.49,130.14) -- (307.49,193.14) -- (292.49,193.14) ;
\draw [line width=1.5]    (169,126) -- (194.5,126) ;
\draw [shift={(198.5,126)}, rotate = 180] [fill={rgb, 255:red, 0; green, 0; blue, 0 }  ][line width=0.08]  [draw opacity=0] (11.61,-5.58) -- (0,0) -- (11.61,5.58) -- cycle    ;
\draw  [line width=1.5]  (197.5,127.37) .. controls (197.5,119.76) and (204.17,113.6) .. (212.4,113.6) .. controls (220.63,113.6) and (227.3,119.76) .. (227.3,127.37) .. controls (227.3,134.97) and (220.63,141.14) .. (212.4,141.14) .. controls (204.17,141.14) and (197.5,134.97) .. (197.5,127.37) -- cycle ;
\draw  [line width=1.5]  (254.17,99.09) -- (292.5,99.09) -- (292.5,157) -- (254.17,157) -- cycle ;
\draw [line width=1.5]    (227.5,128) -- (250.5,128) ;
\draw [shift={(254.5,128)}, rotate = 180] [fill={rgb, 255:red, 0; green, 0; blue, 0 }  ][line width=0.08]  [draw opacity=0] (11.61,-5.58) -- (0,0) -- (11.61,5.58) -- cycle    ;
\draw  [dash pattern={on 4.5pt off 4.5pt}]  (180.97,126.23) -- (181,147) -- (253.5,148) ;
\draw [line width=1.5]    (294.5,130) -- (320.49,130.28) -- (335.5,130.06) ;
\draw [shift={(339.5,130)}, rotate = 539.1600000000001] [fill={rgb, 255:red, 0; green, 0; blue, 0 }  ][line width=0.08]  [draw opacity=0] (11.61,-5.58) -- (0,0) -- (11.61,5.58) -- cycle    ;
\draw  [line width=1.5]  (249.7,194.74) -- (289,174) -- (288.28,214.02) -- cycle ;
\draw    (248.5,164) -- (300.78,89.46) ;
\draw [shift={(302.5,87)}, rotate = 485.04] [fill={rgb, 255:red, 0; green, 0; blue, 0 }  ][line width=0.08]  [draw opacity=0] (8.93,-4.29) -- (0,0) -- (8.93,4.29) -- cycle    ;
\draw [line width=1.5]    (250.56,195.69) -- (212,195) -- (212.37,145.14) ;
\draw [shift={(212.4,141.14)}, rotate = 450.42] [fill={rgb, 255:red, 0; green, 0; blue, 0 }  ][line width=0.08]  [draw opacity=0] (11.61,-5.58) -- (0,0) -- (11.61,5.58) -- cycle    ;
\draw [line width=1.5]    (299.66,57.41) -- (417.5,59) -- (417.41,112.6) ;
\draw [shift={(417.4,116.6)}, rotate = 270.1] [fill={rgb, 255:red, 0; green, 0; blue, 0 }  ][line width=0.08]  [draw opacity=0] (11.61,-5.58) -- (0,0) -- (11.61,5.58) -- cycle    ;
\draw  [line width=1.5]  (49.63,244.45) .. controls (49.63,236.84) and (56.3,230.68) .. (64.53,230.68) .. controls (72.76,230.68) and (79.43,236.84) .. (79.43,244.45) .. controls (79.43,252.06) and (72.76,258.22) .. (64.53,258.22) .. controls (56.3,258.22) and (49.63,252.06) .. (49.63,244.45) -- cycle ;
\draw [line width=1.5]    (64.53,230.68) -- (63.57,143.22) ;
\draw [shift={(63.53,139.22)}, rotate = 449.37] [fill={rgb, 255:red, 0; green, 0; blue, 0 }  ][line width=0.08]  [draw opacity=0] (11.61,-5.58) -- (0,0) -- (11.61,5.58) -- cycle    ;
\draw [line width=1.5]    (8.5,243) -- (47.63,243.41) ;
\draw [shift={(51.63,243.45)}, rotate = 180.6] [fill={rgb, 255:red, 0; green, 0; blue, 0 }  ][line width=0.08]  [draw opacity=0] (11.61,-5.58) -- (0,0) -- (11.61,5.58) -- cycle    ;
\draw  [line width=1.5]  (375.65,128.35) -- (336.61,149.56) -- (336.84,109.54) -- cycle ;
\draw  [line width=1.5]  (167.68,125.85) -- (128.1,146.05) -- (129.37,106.04) -- cycle ;
\draw [line width=1.5]    (78.43,125.45) -- (123.5,125.96) ;
\draw [shift={(127.5,126)}, rotate = 180.64] [fill={rgb, 255:red, 0; green, 0; blue, 0 }  ][line width=0.08]  [draw opacity=0] (11.61,-5.58) -- (0,0) -- (11.61,5.58) -- cycle    ;
\draw  [line width=1.5]  (299.66,57.41) -- (260.54,78.5) -- (260.9,38.47) -- cycle ;
\draw [line width=1.5]    (107.47,125.73) -- (107.5,56) -- (256.5,57.95) ;
\draw [shift={(260.5,58)}, rotate = 180.75] [fill={rgb, 255:red, 0; green, 0; blue, 0 }  ][line width=0.08]  [draw opacity=0] (11.61,-5.58) -- (0,0) -- (11.61,5.58) -- cycle    ;
\draw  [line width=1.5]  (402.5,130.37) .. controls (402.5,122.76) and (409.17,116.6) .. (417.4,116.6) .. controls (425.63,116.6) and (432.3,122.76) .. (432.3,130.37) .. controls (432.3,137.97) and (425.63,144.14) .. (417.4,144.14) .. controls (409.17,144.14) and (402.5,137.97) .. (402.5,130.37) -- cycle ;
\draw [line width=1.5]    (432.3,130.37) -- (457.31,130.8) ;
\draw [shift={(461.31,130.87)}, rotate = 180.99] [fill={rgb, 255:red, 0; green, 0; blue, 0 }  ][line width=0.08]  [draw opacity=0] (11.61,-5.58) -- (0,0) -- (11.61,5.58) -- cycle    ;
\draw [line width=1.5]    (375.65,128.35) -- (400.66,128.78) ;
\draw [shift={(404.66,128.85)}, rotate = 180.99] [fill={rgb, 255:red, 0; green, 0; blue, 0 }  ][line width=0.08]  [draw opacity=0] (11.61,-5.58) -- (0,0) -- (11.61,5.58) -- cycle    ;
\draw  [line width=1.5]  (602.63,128.45) .. controls (602.63,120.84) and (609.3,114.68) .. (617.53,114.68) .. controls (625.76,114.68) and (632.43,120.84) .. (632.43,128.45) .. controls (632.43,136.06) and (625.76,142.22) .. (617.53,142.22) .. controls (609.3,142.22) and (602.63,136.06) .. (602.63,128.45) -- cycle ;
\draw [line width=1.5]    (617.5,62) -- (617.53,110.68) ;
\draw [shift={(617.53,114.68)}, rotate = 269.97] [fill={rgb, 255:red, 0; green, 0; blue, 0 }  ][line width=0.08]  [draw opacity=0] (11.61,-5.58) -- (0,0) -- (11.61,5.58) -- cycle    ;
\draw [line width=1.5]    (632.43,128.45) -- (662.5,128.94) ;
\draw [shift={(666.5,129)}, rotate = 180.92] [fill={rgb, 255:red, 0; green, 0; blue, 0 }  ][line width=0.08]  [draw opacity=0] (11.61,-5.58) -- (0,0) -- (11.61,5.58) -- cycle    ;
\draw (516.5,127.5) node  [scale=1.3,font=\large]  {$\bm{C(z)}$};
\draw (63.53,123.61) node  [scale=1.3,font=\large]  {$\bm{-}$};
\draw (25.84,109.76) node  [scale=1.3,font=\large]  {$\bm{r}$};
\draw (87.35,110.76) node  [scale=1.3,font=\large]  {$\bm{e}$};
\draw (726.84,129.04) node  [scale=1.3,font=\large]  {$\dfrac{\bm{1}}{\bm{ms^{2} +cs+k}}$};
\draw (213.4,125.37) node  [scale=1.3,font=\large]  {$\bm{+}$};
\draw (273.34,128.04) node  [scale=1.3,font=\large]  {$\dfrac{\bm{1}}{\bm{s}}$};
\draw (312,76) node  [scale=1.3,font=\large]  {$\bm{\gamma}$};
\draw (152,284) node   [scale=1.3,align=left] {};
\draw (513,60) node  [scale=1.3,font=\large] [align=left] {{\fontfamily{ptm}\selectfont \textbf{ \ Linear Part of}}\\ \ \ \ \ {\fontfamily{ptm}\selectfont \textbf{Controller}}};
\draw (726,83) node  [scale=1.3,font=\large] [align=left] {{\fontfamily{ptm}\selectfont \textbf{Plant}}};
\draw (64.53,242.61) node  [scale=1.3,font=\large]  {$\bm{+}$};
\draw (25.84,226.76) node  [scale=1.3,font=\large]  {$\bm{n}$};
\draw (273,15) node  [scale=1.3,font=\large] [align=left] {{\fontfamily{ptm}\selectfont \textbf{Reset}}};
\draw (275.84,191.76) node  [scale=1.3,font=\large]  {$\bm{A_{r}}$};
\draw (349.84,126.76) node  [scale=1.3,font=\large]  {$\bm{C_{r}}$};
\draw (141.84,123.76) node  [scale=1.3,font=\large]  {$\bm{B_{r}}$};
\draw (273.84,54.76) node  [scale=1.3,font=\large]  {$\bm{D_{r}}$};
\draw (418.4,128.37) node  [scale=1.3,font=\large]  {$\bm{+}$};
\draw (617.53,126.61) node  [scale=1.3,font=\large]  {$\bm{+}$};
\draw (620,47) node  [scale=1.3,font=\large]  {$\bm{w}$};
\draw (818,116) node [scale=1.3,font=\large]  {$\bm{y}$};
\end{tikzpicture}}
\caption{The block diagram of the whole system for implementing the designed controllers (reset matrices are discretized)}
 \label{F-78}
 \end{figure}

In this part, the time domain results of the designed controllers are compared with each other. To implement controllers (Fig.~\ref{F-78}), each controller is discretized with sample time $T_s=$\SI{100}{\micro\second} using the Tustin method \cite{schmidt2014design,sabatier2015fractional,dastjerdi2018tuning}. Furthermore, to provide the well-posedness property \cite{banos2011reset,guo2015analysis}, there are no reset instants in tandem.    

\begin{figure*}
	\centering
	\begin{subfigure}{0.67\columnwidth}
		\centering
		\includegraphics[width=0.8\hsize]{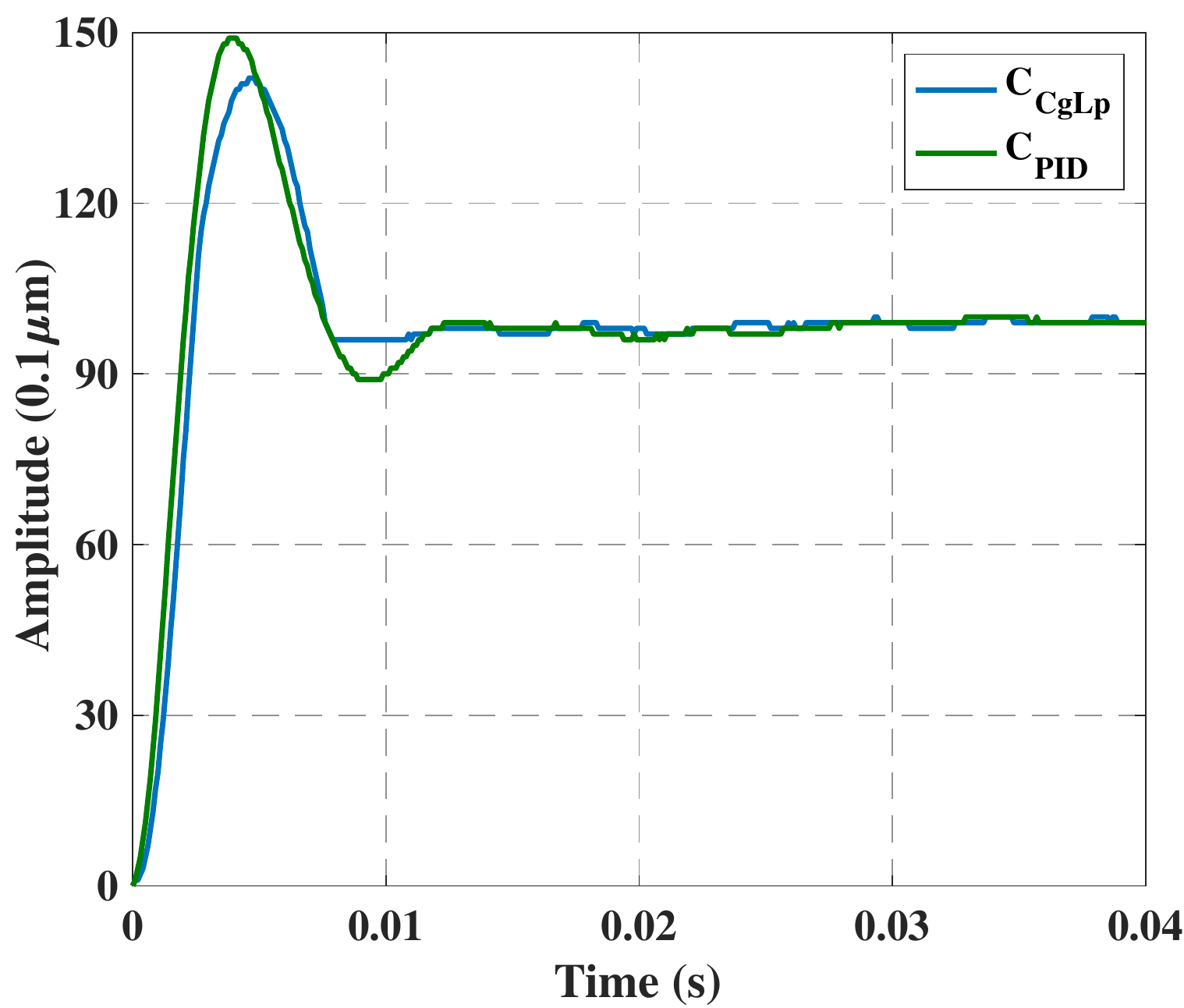}
		  \caption{Step responses of controllers}
    \label{F-08a}
  \end{subfigure}
	\hfil
	\begin{subfigure}{0.67\columnwidth}
		\centering    
		\includegraphics[width=0.8\hsize]{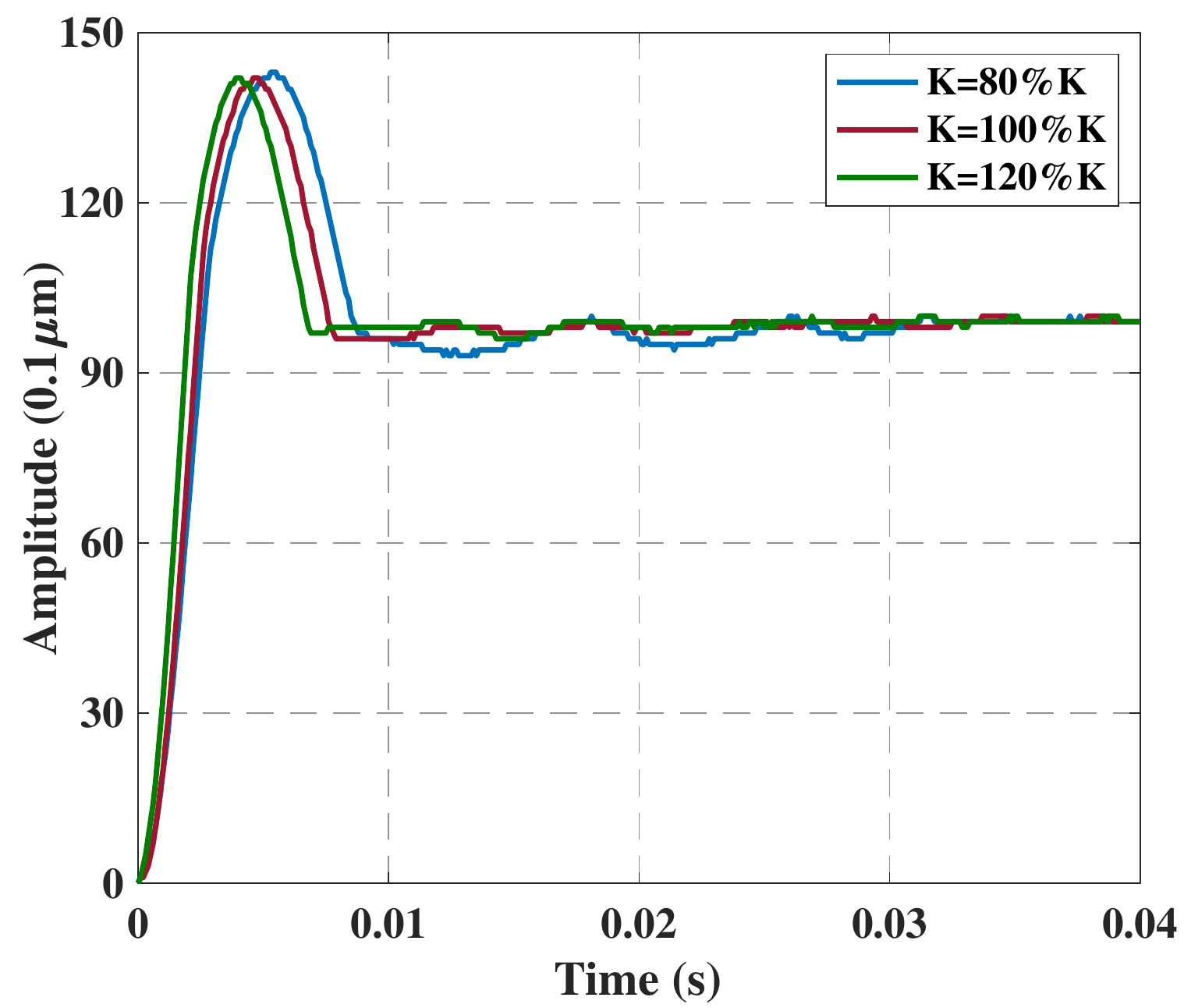}
    \caption{Step responses of $C_{\text{CgLp}}$}
    \label{F-08b}
  \end{subfigure}  
	\hfil
	\begin{subfigure}{0.67\columnwidth}
		\centering
		\includegraphics[width=0.8\hsize]{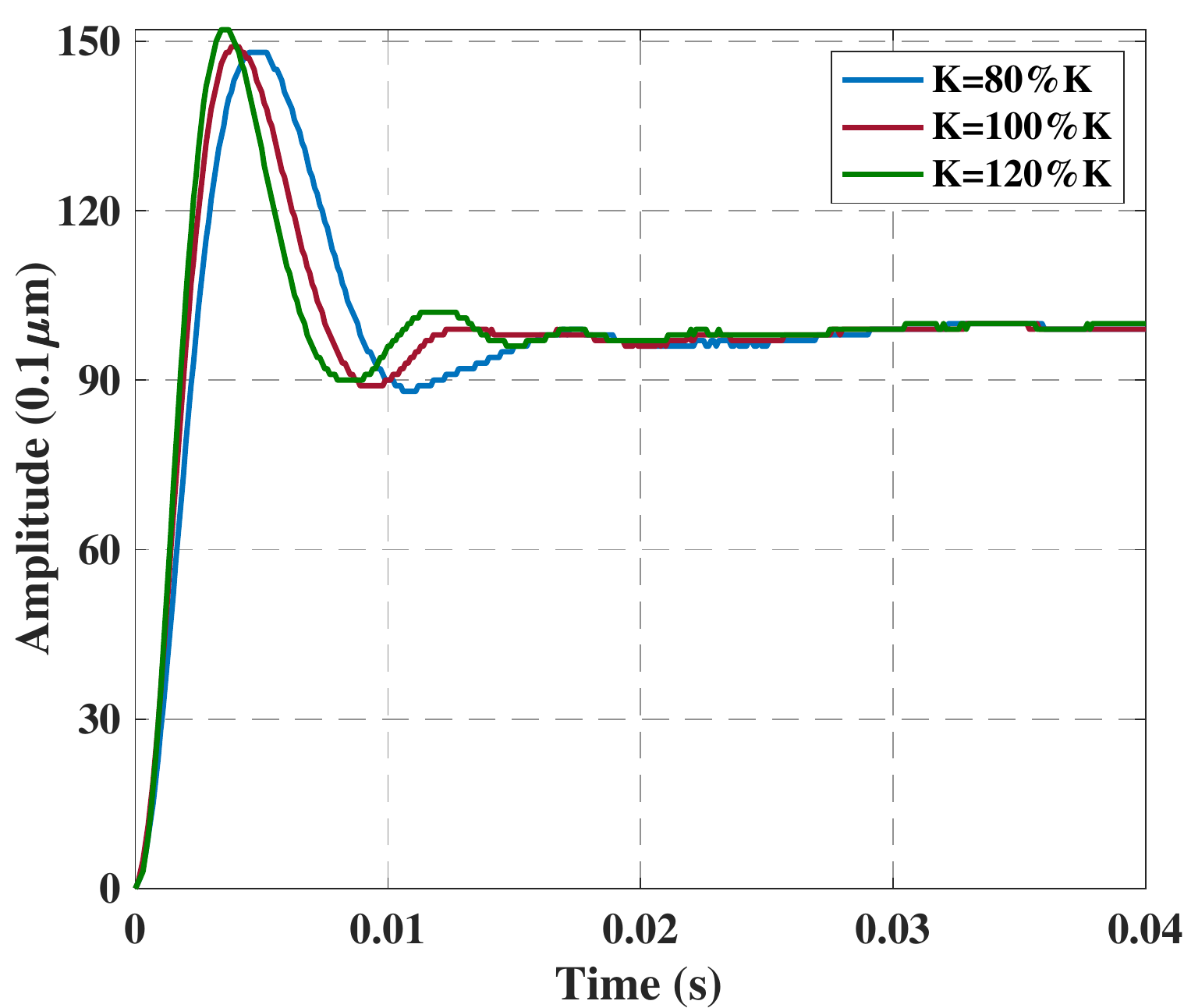}
    \caption{Step responses of $C_{\text{PID}}$}
    \label{F-08d}
  \end{subfigure}  
   \caption{The step responses of controllers with gain variation between $80\%$ to $120\%$ of their nominal values}  
  \label{F-08}
\end{figure*}

The step responses (step of \SI{10}{\micro\meter}) of the system with these controllers are illustrated in Fig.~\ref{F-08}. To assess iso-damping behaviour of the system, the gains of the controllers are varied between $80\%$ to $120\%$ of their nominal values. The step responses have the same rise time while the overshoot of the system with the controller $C_{\text{CgLp}}$ are less than that of the controller $C_{\text{PID}}$ because the modulus margin of the system with the CgLp compensator is less than one with the PID controller. Furthermore, system with the controller $C_{\text{CgLp}}$ has less settling time in comparison with that of with the controller $C_{\text{PID}}$. Besides, step responses of the system with these controllers show iso-damping behaviour indicating. However, $C_{\text{CgLp}}$ provides more robustness against gain variation for the system.

\begin{figure*}
	\centering
	\begin{subfigure}[t]{0.49\columnwidth}
		\centering
		\includegraphics[width=0.955\hsize]{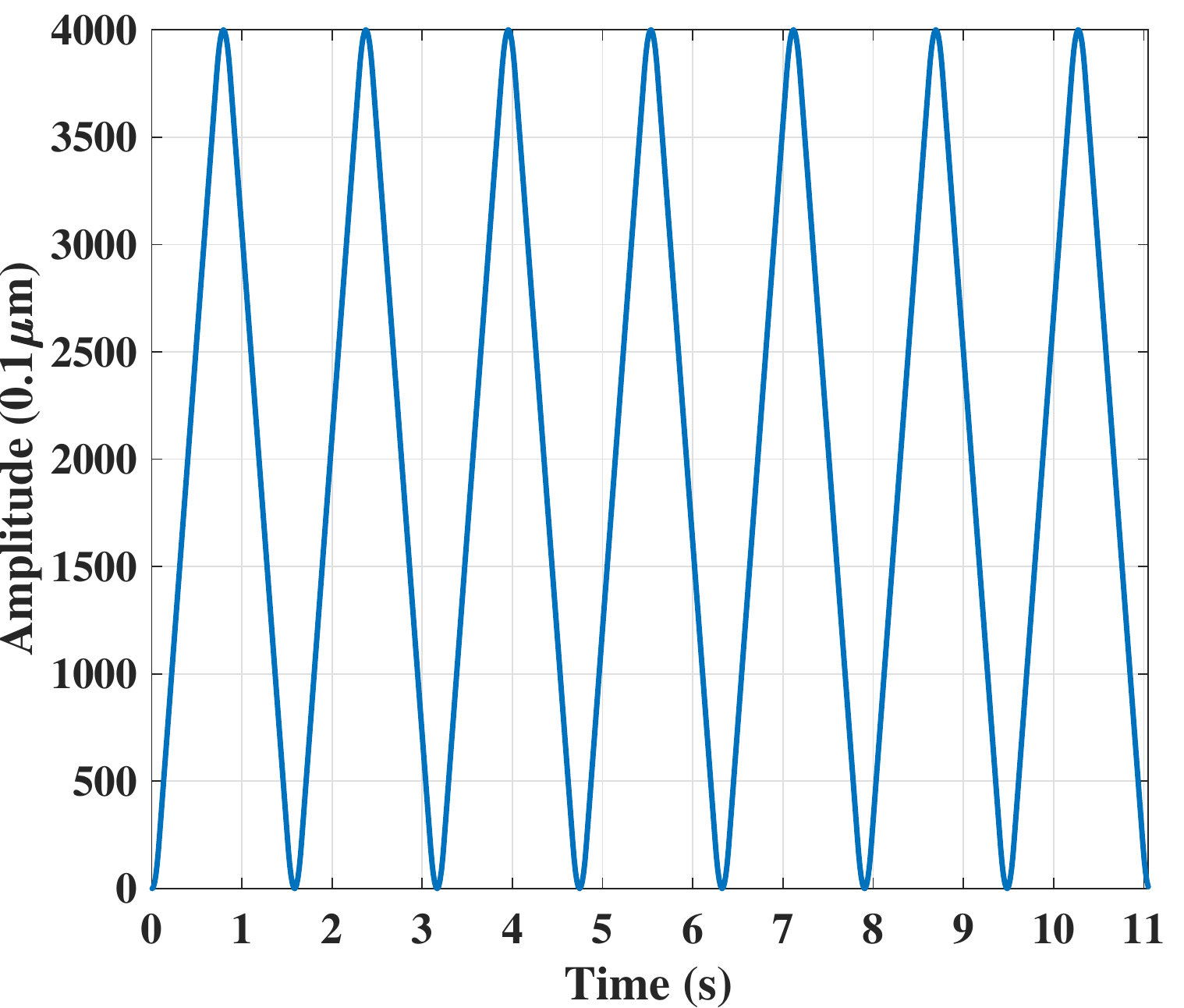}
		  \caption{Triangular reference}
    \label{F-09a}
	\end{subfigure}
	\hfil
	\begin{subfigure}[t]{0.49\columnwidth}
		\centering    
		\includegraphics[width=0.955\hsize]{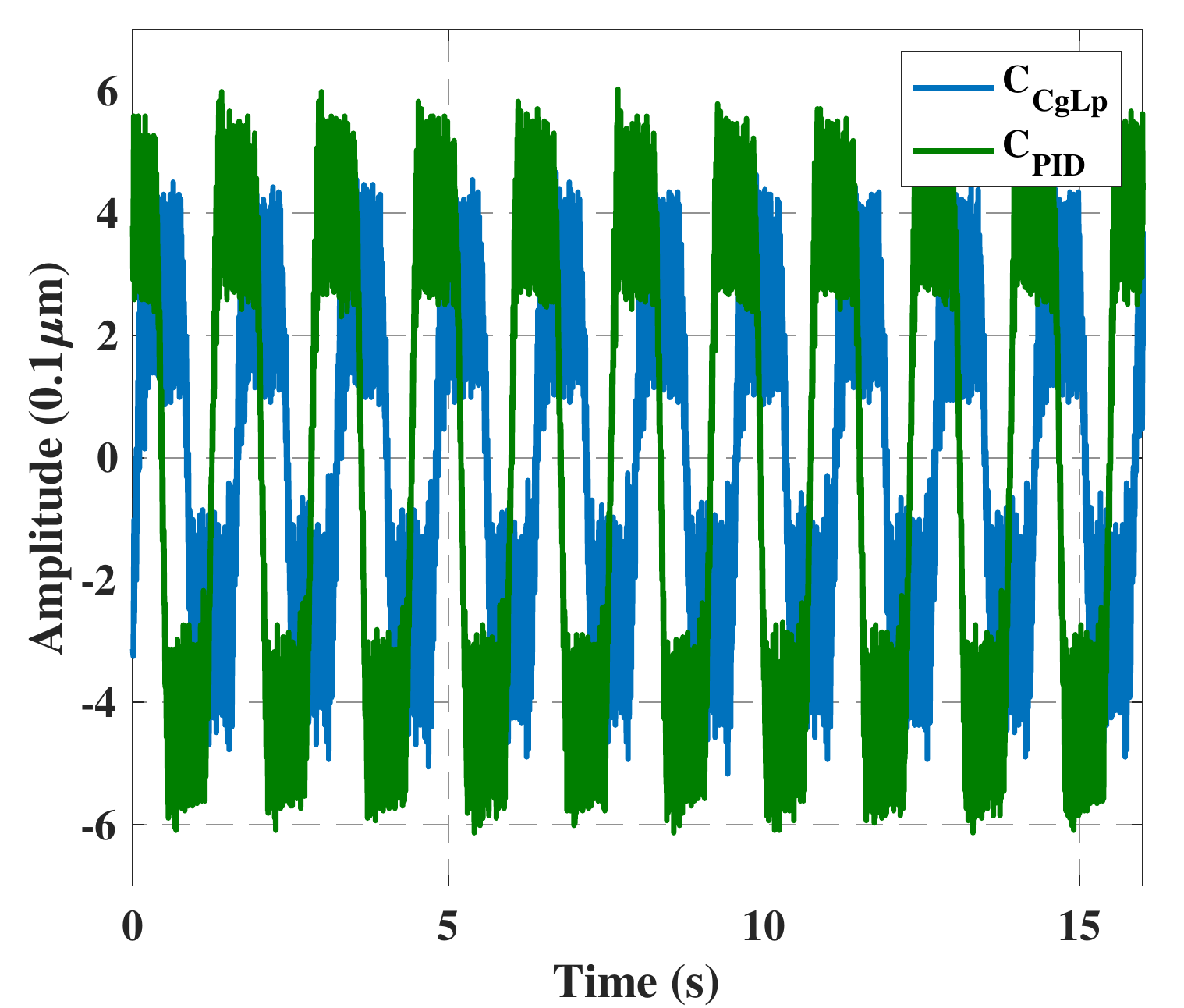}
    \caption{Tracking error of the triangular reference}
    \label{F-09c}
	\end{subfigure}
	\hfil
	\begin{subfigure}[t]{0.49\columnwidth}
		\centering
		\includegraphics[width=0.955\hsize]{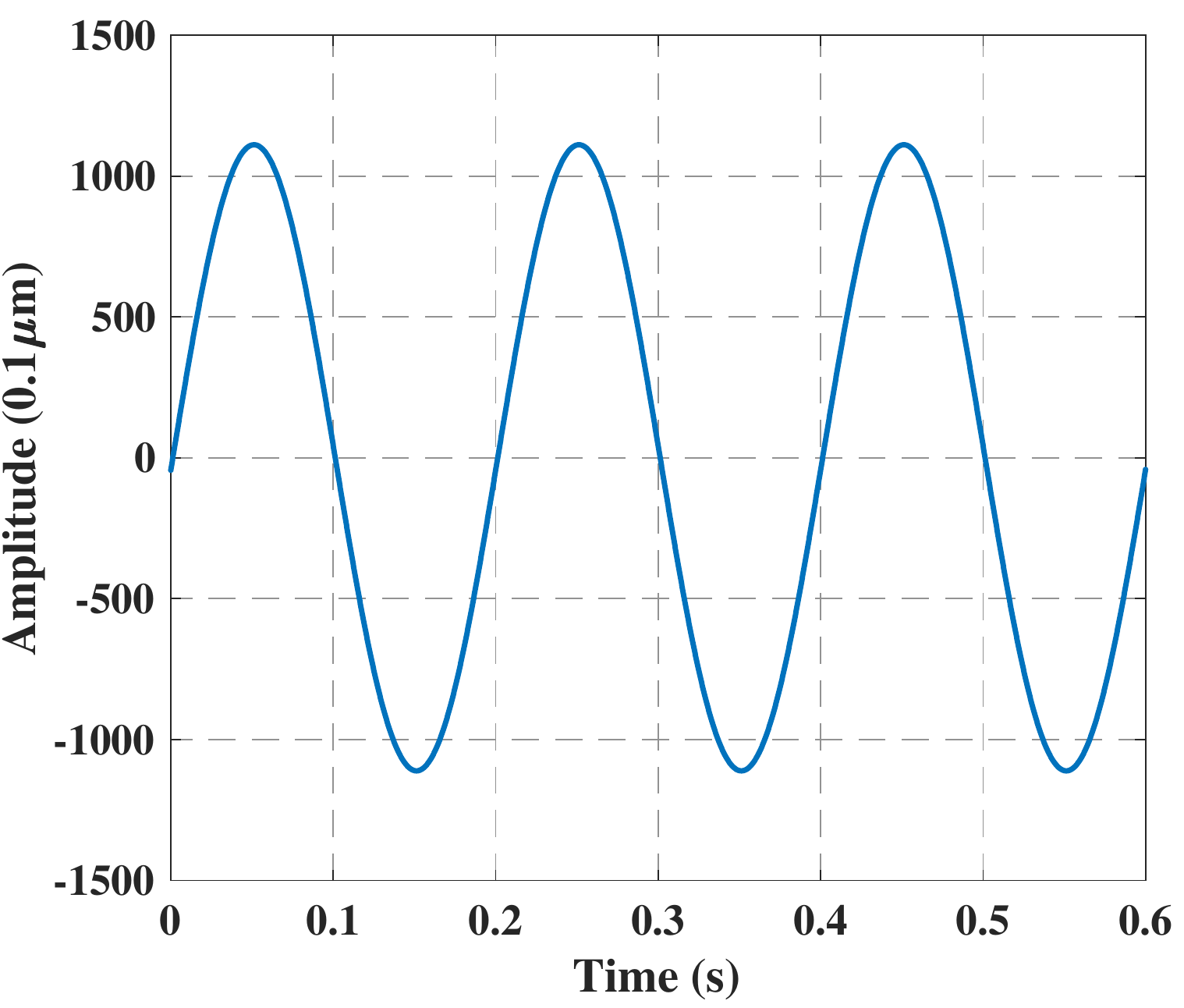}
    \caption{Sinusoidal reference}
    \label{F-09b}
	\end{subfigure}
	\hfil
	\begin{subfigure}[t]{0.49\columnwidth}
		\centering 
		\includegraphics[width=0.955\hsize]{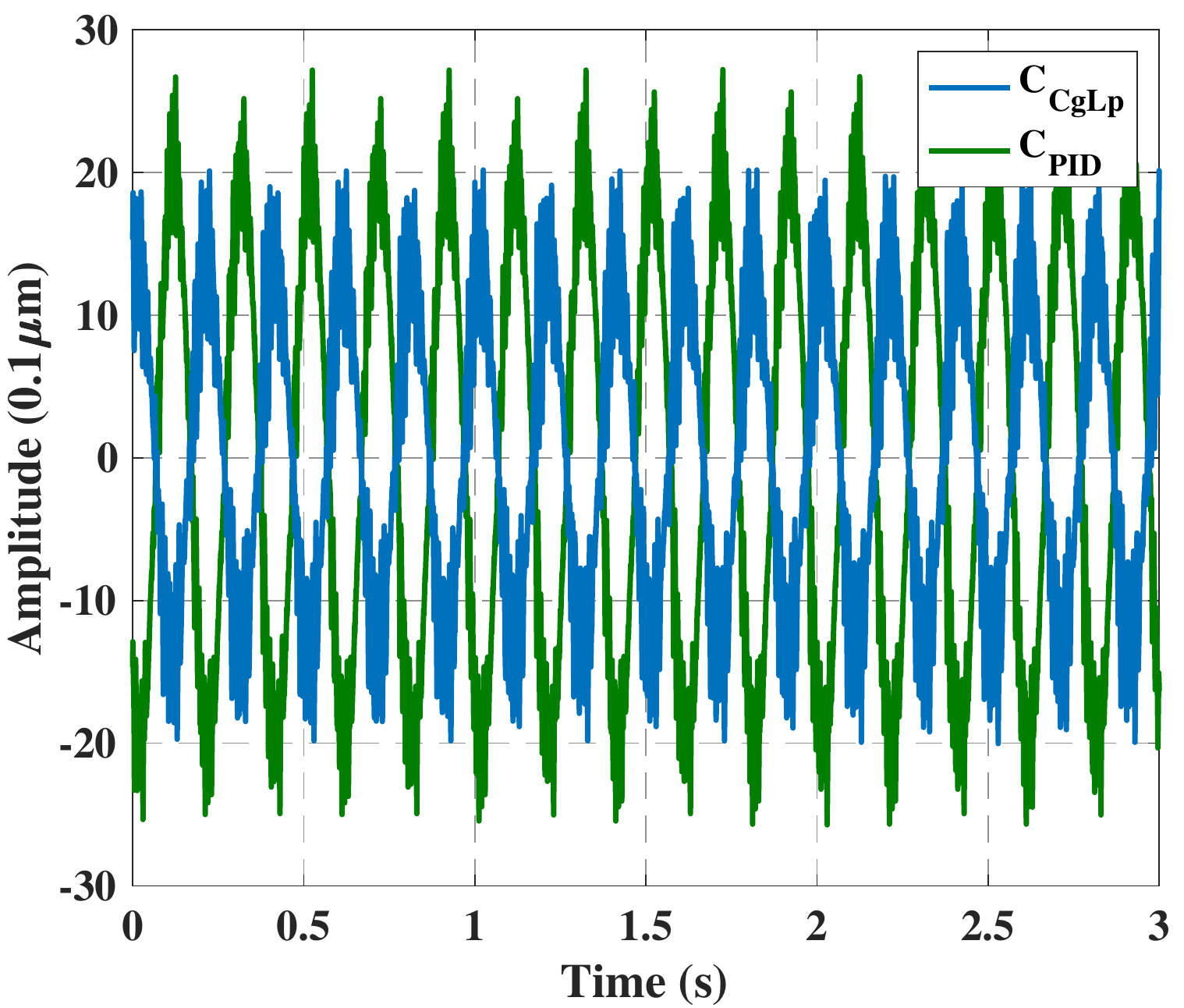}
    \caption{Tracking error of the sinusoidal reference}
    \label{F-09d}
	\end{subfigure}
	\caption{Tracking performance of the designed controllers for a triangular and a sinusoidal references}
	\label{F-09}
\end{figure*}

In order to compare tracking performances of the systems with both controllers, one triangular reference with the amplitude of \SI{400}{\micro\meter} (Fig.~\ref{F-09a}) and one sinusoidal reference $r(t)=111\sin(10\pi t)$ \SI{}{\micro\meter} (Fig.~\ref{F-09b}) are applied to the system. As was predicted by $S_\infty$ (Fig.~\ref{F-06b}), the system with the controller $C_{\text{CgLp}}$ has a better performance at $5$Hz (Fig.~\ref{F-09d}). In addition, $S_\infty$ (Fig.~\ref{F-06b}) precisely predicts the maximum error of the system with the controller $C_{\text{CgLp}}$ for the sinusoidal reference. Note that, for the sake of brevity, we only show the result at $5$Hz while the tracking performance of the system with the controller $C_{\text{CgLp}}$ is better than one with the controller $C_{\text{PID}}$ for the sinusoidal reference for all frequencies less than $10 Hz$. As shown in Fig.~\ref{F-09c}, the system with the controller $C_{\text{CgLp}}$ also has the better tracking performance than that of the controller $C_{\text{PID}}$ for the triangular reference (Fig.~\ref{F-09a}) which is a combination of several frequencies. For these trajectories, the tracking performance of the system is improved by 30\% using the controller $C_{\text{CgLp}}$.  

\begin{figure}[thbp]
  \centering
 \begin{subfigure}{0.49\linewidth}
    \centering
\includegraphics[width=\linewidth]{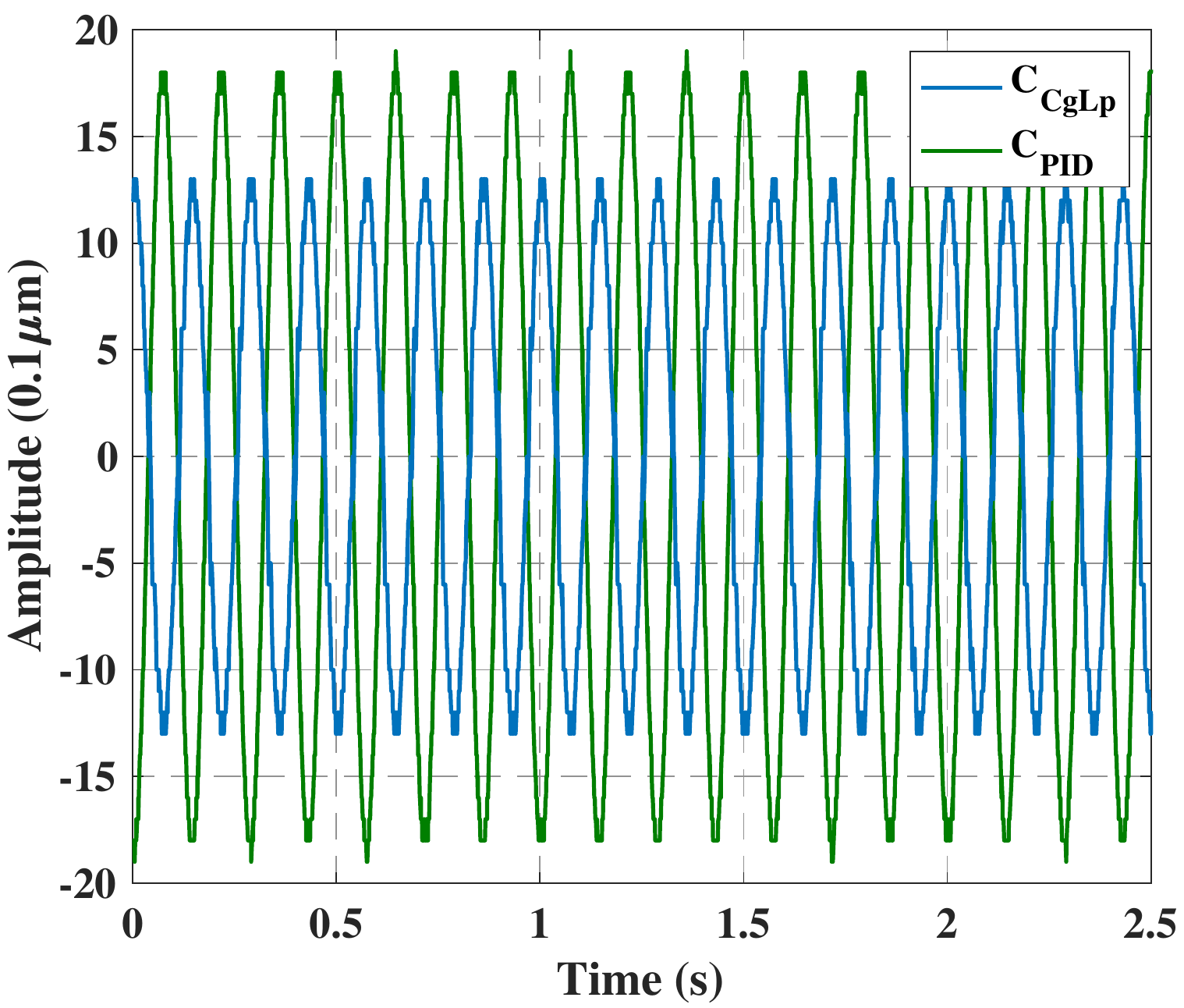}
    \caption{Disturbance rejection}
    \label{F-10a}
  \end{subfigure}
\begin{subfigure}{0.49\linewidth}
    \centering
\includegraphics[width=\linewidth]{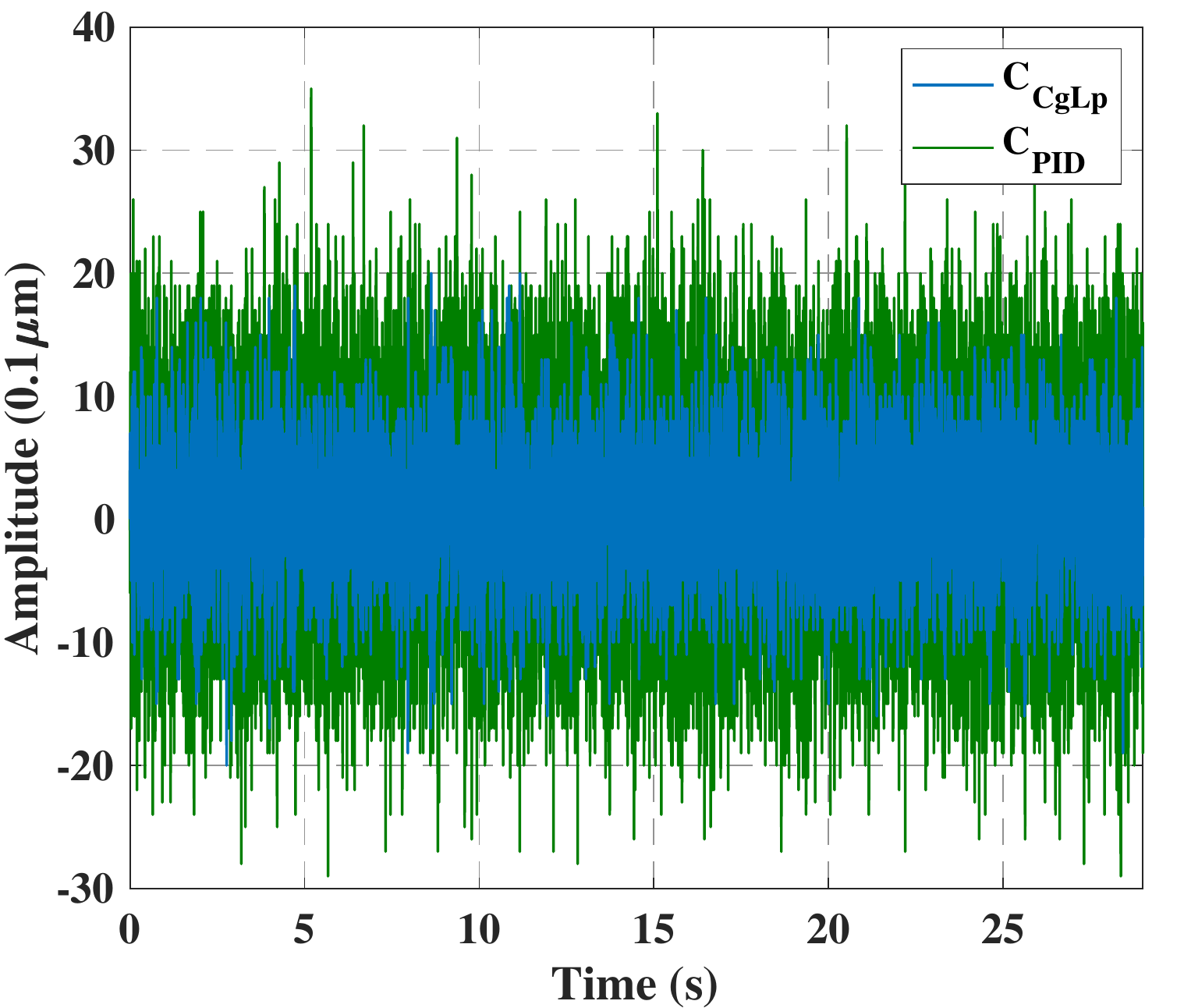}
    \caption{Noise rejection}
    \label{F-10b}
  \end{subfigure}  
   \caption{Disturbance and noise rejection capabilities of the system with the designed controllers}  
  \label{F-10}
\end{figure} 
Figure \ref{F-10} compares the noise and disturbance rejection capabilities of the system with the designed controllers. To study the noise rejection capabilities of the system with the controllers, a white noise with a maximum amplitude of $\SI{5}{\micro\meter}$ is applied to the system. As was expected from $T_\infty$ (Fig.~\ref{F-06a}), the noise rejection capability of the system with the controller $C_{\text{CgLp}}$ is better than one with the controller $C_{\text{PID}}$ (Fig.~\ref{F-10b}). It can be said that using $C_{\text{CgLp}}$ enhance the noise rejection capability of the system by 40\%. In order to evaluate the abilities of the system with the designed controllers for attenuating disturbances, a sinusoidal disturbance $w(t)=190\sin(14\pi t)$ \SI{}{\micro\ampere} is applied to the system. Similar to the $PS_\infty$ prediction (Fig.~\ref{F-07a}), the system with the controller $C_{\text{CgLp}}$ has the optimal disturbance rejection performance (Fig.~\ref{F-10a}). The disturbance rejection capability of the system is improved by 30\% using the controller $C_{\text{CgLp}}$. 

To wrap up, the system with the tuned CgLp compensator has less overshoot, the same rise time, better tracking performance for frequencies less than 10 Hz, less modulus margin, better noise and disturbance rejection capabilities than those of the system with the controller PID. 
\section{Conclusion}\label{sec:4}
This paper has proposed a frequency-domain tuning method for CgLp compensators based on the defined pseudo-sensitivities for reset control systems. In this method, a PID+CgLp structure is considered, and its parameters are tuned such that the pseudo-sensitivity is minimized under several constraints. Also, the tuned CgLp compensator with this method, makes the system robust against gain variations. To show the effectiveness of the proposed approach, the performance of this tuned CgLp is compared with a linear PID. The results show that the frequency framework is reliable for tuning CgLp compensators. Furthermore, the tuned CgLp can achieve more favourable dynamic performance than the PID controller for the precision motion stage. The tracking performance, the disturbance rejection capability, and the noise rejection capability of the system are improved by 30\% using the CgLp compensator. Indeed, this method, which allows for tuning in the frequency-domain, opens doors for the implementation of reset controllers in industrial applications.

\bibliographystyle{IEEEtran}
\bibliography{phd}

\begin{thebibliography}{10}
\providecommand{\url}[1]{#1}
\csname url@samestyle\endcsname
\providecommand{\newblock}{\relax}
\providecommand{\bibinfo}[2]{#2}
\providecommand{\BIBentrySTDinterwordspacing}{\spaceskip=0pt\relax}
\providecommand{\BIBentryALTinterwordstretchfactor}{4}
\providecommand{\BIBentryALTinterwordspacing}{\spaceskip=\fontdimen2\font plus
\BIBentryALTinterwordstretchfactor\fontdimen3\font minus
  \fontdimen4\font\relax}
\providecommand{\BIBforeignlanguage}[2]{{%
\expandafter\ifx\csname l@#1\endcsname\relax
\typeout{** WARNING: IEEEtran.bst: No hyphenation pattern has been}%
\typeout{** loaded for the language `#1'. Using the pattern for}%
\typeout{** the default language instead.}%
\else
\language=\csname l@#1\endcsname
\fi
#2}}
\providecommand{\BIBdecl}{\relax}
\BIBdecl

\bibitem{dastjerdi2019linear}
A.~A. Dastjerdi, B.~M. Vinagre, Y.~Chen, and S.~H. HosseinNia, ``Linear
  fractional order controllers; a survey in the frequency domain,''
  \emph{Annual Reviews in Control}, 2019.

\bibitem{horowitz1975non}
I.~Horowitz and P.~Rosenbaum, ``Non-linear design for cost of feedback
  reduction in systems with large parameter uncertainty,'' \emph{International
  Journal of Control}, vol.~21, no.~6, pp. 977--1001, 1975.

\bibitem{guo2009frequency}
Y.~Guo, Y.~Wang, and L.~Xie, ``Frequency-domain properties of reset systems
  with application in hard-disk-drive systems,'' \emph{IEEE Transactions on
  Control Systems Technology}, vol.~17, no.~6, pp. 1446--1453, 2009.

\bibitem{clegg1958nonlinear}
J.~C. {Clegg}, ``A nonlinear integrator for servomechanisms,''
  \emph{Transactions of the American Institute of Electrical Engineers, Part
  II: Applications and Industry}, vol.~77, no.~1, pp. 41--42, 1958.

\bibitem{beker2004fundamental}
O.~Beker, C.~Hollot, Y.~Chait, and H.~Han, ``Fundamental properties of reset
  control systems,'' \emph{Automatica}, vol.~40, no.~6, pp. 905 -- 915, 2004.

\bibitem{hazeleger2016second}
L.~{Hazeleger}, M.~{Heertjes}, and H.~{Nijmeijer}, ``Second-order reset
  elements for stage control design,'' in \emph{American Control Conference
  (ACC)}, 2016, pp. 2643--2648.

\bibitem{guo2015analysis}
Y.~Guo, L.~Xie, and Y.~Wang, \emph{Analysis and Design of Reset Control
  Systems}.\hskip 1em plus 0.5em minus 0.4em\relax Institution of Engineering
  and Technology, 2015.

\bibitem{saikumar2019constant}
N.~{Saikumar}, R.~K. {Sinha}, and S.~H. {HosseinNia}, ```{C}onstant in gain
  {L}ead in phase' element-application in precision motion control,''
  \emph{IEEE/ASME Transactions on Mechatronics}, vol.~24, no.~3, pp.
  1176--1185, 2019.

\bibitem{barreiro2014reset}
A.~Barreiro, A.~Ba{\~n}os, S.~Dormido, and J.~A. Gonz{\'a}lez-Prieto, ``Reset
  control systems with reset band: Well-posedness, limit cycles and stability
  analysis,'' \emph{Systems \& Control Letters}, vol.~63, pp. 1--11, 2014.

\bibitem{banos2014tuning}
A.~Ba{\~n}os and M.~A. Dav{\'o}, ``Tuning of reset proportional integral
  compensators with a variable reset ratio and reset band,'' \emph{IET Control
  Theory \& Applications}, vol.~8, no.~17, pp. 1949--1962, 2014.

\bibitem{zheng2007improved}
J.~Zheng, Y.~Guo, M.~Fu, Y.~Wang, and L.~Xie, ``Improved reset control design
  for a pzt positioning stage,'' \emph{IEEE International Conference on
  Control Applications}.\hskip 1em plus 0.5em minus 0.4em\relax IEEE, 2007, pp.
  1272--1277.

\bibitem{banos2011reset}
A.~Ba{\~n}os and A.~Barreiro, \emph{Reset control systems}.\hskip 1em plus
  0.5em minus 0.4em\relax Springer Science $\&$ Business Media, 2011.

\bibitem{van2018hybrid}
S.~J. A.~M. {Van den Eijnden}, Y.~{Knops}, and M.~F. {Heertjes}, ``A hybrid
  integrator-gain based low-pass filter for nonlinear motion control,''
  \emph{IEEE Conference on Control Technology and Applications (CCTA)}, 2018,
  pp. 1108--1113.

\bibitem{valerio2019reset}
D.~Val{\'e}rio, N.~Saikumar, A.~A. Dastjerdi, N.~Karbasizadeh, and S.~H.
  HosseinNia, ``Reset control approximates complex order transfer functions,''
  \emph{Nonlinear Dynamics}, pp. 1--15, 2019.

\bibitem{Houu}
H.~Xiaojun, A.~Ahmadi~Dastjerdi, N.~Saikumar, and S.~HosseinNia, ``Tuning of
  `{C}onstant in gain {L}ead in phase ({CgLp})' reset controller using
  {H}igher-{O}rder {S}inusoidal {I}nput {D}escribing {F}unction ({HOSIDF}),'' \emph{Australian and New Zealand Control Conference (ANZCC)}, 2020.

\bibitem{Mahmoud}
M.~Shirdast~Bahnamiri, N.~Karbasizadeh, A.~Ahmadi~Dastjerdi, N.~Saikumar, and
  S.~HosseinNia, ``Tuning of {CgLp} based reset controllers: Application in
  precision positioning systems,'' \emph{IFAC World Congress}, 2020.

\bibitem{ALIAUTO}
A.~{Ahmadi Dastjerdi}, N.~{Saikumar}, D.~{Valerio}, and S.~{Hassan HosseinNia},
  ``{Closed-loop frequency analyses of reset systems},'' \emph{arXiv e-prints},
  p. arXiv:2001.10487, Jan 2020.

\bibitem{AliCDC}
A.~{A. Dastjerdi}, A.~{Astolfi}, and S.~H. {HosseinNia}, ``A frequency-domain
  stability method for reset systems,'' \emph{IEEE 59th Conference on
  Decision and Control}, 2020.

\bibitem{saikumar2019complex}
``Complex order control for improved loop-shaping in precision positioning,''
  in \emph{IEEE 58th Conference on Decision and Control}, 2019, pp. 7956--7962.

\bibitem{polenkova2012stability}
S.~Polenkova, J.~W. Polderman, and R.~Langerak, ``Stability of reset systems,'' \emph{Proceedings of the 20th International Symposium on Mathematical
  Theory of Networks and Systems}, 2012, pp. 9--13.

\bibitem{schmidt2014design}
R.~M. Schmidt, G.~Schitter, and A.~Rankers, \emph{The Design of High
  Performance Mechatronics High-Tech Functionality by Multidisciplinary System
  Integration}.\hskip 1em plus 0.5em minus 0.4em\relax IOS Press, 2014.

\bibitem{Toolbox}
\BIBentryALTinterwordspacing
A.~A. Dastjerdi. Toolbox for frequency analysis of reset control systems.
  [Online]. Available:\\
  \url{https://www.tudelft.nl/en/3me/about/departments/precision-and-microsystems-engineering-pme/research/mechatronic-system-design-msd/msd-research/motion-control/toolbox-frequency-analysis-of-reset-control-systems/}
\BIBentrySTDinterwordspacing

\bibitem{Caipaper}
C.~Cai, A.~A. Dastjerdi, N.~Saikumar, and S.~HosseinNia, ``The optimal sequence
  for reset controllers,'' \emph{$18^{th}$ European Control Conference (ECC}, 2020.

\bibitem{dastjerdi2018tuning}
A.~A. Dastjerdi, N.~Saikumar, and S.~H. HosseinNia, ``Tuning guidelines for
  fractional order {PID} controllers: Rules of thumb,'' \emph{Mechatronics},
  vol.~56, pp. 26 -- 36, 2018.

\bibitem{de2016novel}
R.~De~Keyser, C.~I. Muresan, and C.~M. Ionescu, ``A novel auto-tuning method
  for fractional order {PI/PD} controllers,'' \emph{ISA transactions}, vol.~62,
  pp. 268--275, 2016.

\bibitem{sabatier2015fractional}
J.~Sabatier, P.~Lanusse, P.~Melchior, and A.~Oustaloup, \emph{Fractional order
  differentiation and robust control design}.\hskip 1em plus 0.5em minus
  0.4em\relax Springer, 2015, vol.~77.

\bibitem{al2000approximation}
S.~Al-Amer and F.~Al-Sunni, ``Approximation of time-delay systems,'' \emph{Proceedings of the American Control Conference. ACC (IEEE Cat. No.
  00CH36334)}, vol.~4.\hskip 1em plus 0.5em minus 0.4em\relax IEEE, 2000, pp.
  2491--2495.

\end{thebibliography}
\end{document}